\documentclass[%
reprint,
amsmath,amssymb,
aps,
pra,
]{revtex4-2}
\usepackage{hyperref}
\usepackage{physics}
\usepackage{xcolor}
\usepackage[caption=false]{subfig}
\usepackage{graphicx}% Include figure files
\usepackage{dcolumn}% Align table columns on decimal point
\usepackage{bm}% bold math
\begin{document}

% \preprint{APS/123-QED}

\title{ Associative Memory for Quantum Entangled States}
% \thanks{A footnote to the article title}

\author{Qiyuan Hu}
\affiliation{
Department of Physics, The University of Chicago, Chicago, Illinois 60637, USA
}

\author{Kartiek Agarwal}
\affiliation{
Materials Science Division, Argonne National Laboratory, Lemont, Illinois 60439, USA
}
\author{Ivar Martin}
\affiliation{
Department of Physics, The University of Chicago, Chicago, Illinois 60637, USA
}
\affiliation{
Materials Science Division, Argonne National Laboratory, Lemont, Illinois 60439, USA
}

% \altaffiliation[Also at ]{Physics Department, XYZ University.}
% \collaboration{Collaboration}

\date{\today}

\begin{abstract}
We generalize the canonical Hopfield model of associative memory, which stores classical states of individual spins, to store entangled quantum states.
The construction is based on truncated projector-sum Hamiltonian, and we explicitly consider a special case that stores dimer singlet coverings of a spin-1/2 system. We analyze the stability of the stored memories and find that the capacity -- the number of stable patterns --  scales faster  with the system size than one would expect from a naive analogy with the standard Hopfield model, suggesting a possible quantum advantage. 

% \begin{description}
% Secondary publications and information retrieval purposes.
% \end{description}
\end{abstract}

%\keywords{Suggested keywords}%Use showkeys class option if keyword display desired

\maketitle

%\tableofcontents

\section{\label{sec:introduction}Introduction}
The Hopfield network~\cite{hopfield82} is a model of associative memory, in which a spin system stores a collection of binary patterns and subsequently retrieves a stored pattern from a partial or corrupted input by means of iterative energy minimization. In this model, memory patterns are attractors in an energy landscape. They are imprinted by a choice of pairwise interactions between spins, such as the Hebbian learning rule \cite{hebb1949}. %Retrieval occurs through dynamical evolution of the system toward local energy minima corresponding to the patterns. 

The principal figure of merit in  associative memory is its storage capacity --  the maximum number of patterns that can be stored before interference between them destroys the retrieval of individual memories. Using mean field theory, replica trick, or statistical methods such as signal-to-noise analysis, one finds that in the thermodynamic limit retrieval of random uncorrelated patterns in the Hopfield model is stable up to a memory load that scales linearly with system size \cite{hopfield82,amit1986,amit1987}. Beyond this threshold, the system undergoes a transition to a spin glass phase dominated by frustration and spurious minima, in which retrieval of individual memories is no longer possible.

Several generalizations of the original Hopfield model have been proposed to overcome its linear storage limitation. Of particular note is the extension to models that include simultaneous interactions of  more than two spins \cite{arenzon1993,baldi1987,bovier2001,hopfield2016}. Encoding memories through multi-spin couplings can significantly alter the structure of the energy landscape and lead to superlinear or even exponential capacities \cite{gardner1987,vermet2017,bao2022,mezard2024}. These constructions are known as dense associative memories or modern Hopfield networks.

More recently, there has been an emergent interest in quantum generalizations of associative memory models \cite{ventura2000, lewenstein2021, marsh2025, schuld2014, lloyd2018, meinhardt2020, nishimori1996, ma1993, nishimori2015, okajima2025, richard2026, rotondo2018, fiorelli2022, fiorelli2023, torres2024, kimura2025}, motivated by the possibility that quantum mechanical effects may provide mechanisms for enhancing storage capacity or give rise to qualitatively new  behaviors. One line of research seeks to realize associative memory in isolated quantum systems by embedding Hopfield-like memories into pure quantum states, thereby leveraging quantum parallelism to improve capacity or reduce computational complexity \cite{schuld2014, lloyd2018, meinhardt2020}. Specifically, exponentially many classical patterns can be encoded using resources that scale only polynomially with the system size \cite{lloyd2018}. Larger storage capacity compared with the classical Hopfield model has also been found in numerical simulations of small systems, though whether this advantage persists for larger systems is yet to be confirmed \cite{meinhardt2020}.
% \ivar{need to say here whether this approach is successful} 
%\textcolor{red}{You should remove this line as it does not really correspond at all to an approach trying to increase capacity}. 

Quantum effects can also be incorporated by replacing classical Ising spins with fully quantum ``vector" spins, described by all three Pauli matrices. The simplest way to include quantum dynamics by turning on a transverse magnetic field in the Hopfield network, unfortunately, only suppresses the capacity by introducing quantum fluctuations, which play a role qualitatively similar to thermal fluctuations  \cite{nishimori1996, ma1993, nishimori2015, okajima2025}.
Quantumness needs not be always detrimental, however: allowing more complex non-Ising interactions of quantum spins in fact can enhance storage capacity, as was shown in \cite{richard2026} by a quantum generalization of the replica trick. 
Direct verification of the enhancement, however, remains challenging due to the small system sizes amenable to controlled quantum simulation. 

Yet another line of research goes beyond the treatment of the Hopfield network as an isolated system and instead considers it as an open quantum system governed by the Lindblad dissipative dynamics \cite{rotondo2018, fiorelli2022, fiorelli2023, torres2024, kimura2025}. In this framework, memory retrieval emerges from the interplay between quantum coherent evolution and engineered dissipation, producing non-classical non-equilibrium behavior \cite{rotondo2018, torres2024, kimura2025} and modifying the storage and retrieval properties of the network \cite{fiorelli2022, fiorelli2023}.

Despite their differences, all studies mentioned above share a common feature: the stored memories are assumed to be unentangled (classical) patterns, i.e. product states of  either binary variables or continuous three-dimensional unit vectors that describe states of individual spins. Quantum effects enter through the dynamics, the implementation of the network, or the retrieval protocol, while the information being stored remains purely classical. This obvious limitation begs a question: can associative memory models be extended to the storage of intrinsically quantum entangled states?

In this work, we answer this question in the affirmative. The specific example that we consider is when  each memory is represented by a perfect matching of $N$ spins into pairs, with each pair forming a spin singlet. The individual memory states are then valence-bond solid states, characterized by distinct patterns of quantum entanglement rather than classical spin configurations. In fact, the expectation value of every individual spin in such a state identically vanishes. To encode a memory as a low energy state of a Hamiltonian, we propose a general truncated projector construction. When applied to  the simple classical product states, this approach recovers the sequence of dense associative memory models mentioned above. When applied to the dimer covering states, it produces a sequence of spin-models with the capacity governed by the truncation order in the dimer-dimer interactions. 
Furthermore, not only are these memory states fundamentally different from the spin product states, but the storage capacity also scales with a higher power of the system size than that of the classical Hopfield counterparts with interactions of the same order.  Thus, even moderate entanglement built into the valence bond solid states enables a qualitatively different associative memory model with higher capacity.

%Motivated by the Hopfield construction, we introduce a Hamiltonian that energetically favors the singlet bonds associated with a given set of dimer coverings and investigate how the stability of these memories evolves as the memory load is increased.

\section{Method and Model}
\label{section:method_model}
In this section, we present a general method for constructing low-operator-weight (`$k$-local') Hamiltonians that encode a set of quantum mechanical states as their approximate ground states. We first illustrate how this approach reproduces the classical dense associative memory models, illuminating the origin of their increased up-to-exponential capacity. After that, we show  how it can be applied to construct models that  store quantum entangled states.

\subsection{Method}
\label{section:method}
Suppose we want to encode a set of orthogonal quantum states $\ket{\psi^\mu}$ as the ground states of some Hamiltonian. 
Formally, this is very simple to accomplish via projectors,
\begin{eqnarray}
    H = -\sum_{\mu= 1}^K \ket{\psi^\mu}\bra{\psi^\mu}.\label{eq:chop_h_proj0}
\end{eqnarray}
This guarantees that the desired states occupy a $K$-fold degenerate ground state subspace with eigenenergy $-1$, while all the other states reside in the null space of the Hamiltonian.  In fact, it is possible to encode any number of states, up to the Hilbert space dimension, as eigenstates with energy $-1$. The problem with this construction, even if we want to store simple product states, is that the Hamiltonian will contain large-weight (up to the system size) operators, making it unwieldy. Also, the ground state manifold is degenerate, so any linear combination of memories will also be a memory.  Therefore, if our goal is to create a model that recalls a specific stored memory, we cannot simply diagonalize the Hamiltonian and pick an arbitrary low-lying eigenstate; rather, we would need to supply additional information about the memory states, such as their degree of entanglement \footnote{Alternatively, by randomizing the weights of the projectors the energetic degeneracy of the memory eigenstates can be lifted, making the memories into unique eigenstates.}.

The issues of both high operator weight and ground-state degeneracy can be resolved by truncating the projectors at some order in elemental local operators. The price to pay for this truncation is that the stored patterns will no longer be exact eigenstates, but the degeneracy of the low energy spectrum will generally be lifted. As the truncation order increases, the storage capacity can approach the exponential limit set by the Hilbert-space dimension.

Next, we demonstrate application of this method  to the spin product states, where it reproduces the well-known dense associative memory models, and then how it  can be used to construct new class of models  that store
the simplest entangled states comprised of non-overlapping singlets.

\subsection{Classical Hopfield Construction}
\label{section:classical_hopfield_model}
The memories of a classical Hopfield network are classical bit strings, with each bit indicating whether a particular spin is pointing up or down. We can encode these states quantum mechanically as product states of $N$ spins, $\ket{\psi^\mu_c} = \otimes_i\ket{\xi_i^\mu }$, where $\xi^\mu_i = \pm 1$, $i = 1, \dots, N$ and $\mu = 1, \dots, K$. Given a set of $K$ stored memories, we can define the Hamiltonian as a sum of projectors onto the corresponding product states,
\begin{equation}
\label{eq:chop_h_proj}
    H^c = - \sum_{\mu=1}^K \ket{\psi^\mu_c}\bra{\psi^\mu_c} = -\sum_{\mu=1}^K \prod_{i=1}^N \frac{1+\xi^\mu_i \sigma_i^z}{2}.
\end{equation}
Expanding the product of single-spin projectors generates interaction terms of increasing order in $\sigma^z$. The first few contributions are
\begin{subequations}
\label{eq:chop_h_expand}
\begin{align}
\begin{split}
\label{eq:chop_h1}
  H_1^c =& -\frac{1}{2^N} \sum_\mu \sum_i \xi^\mu_i \sigma_i^z,
\end{split}\\
\begin{split}
\label{eq:chop_h2}
     H_2^c =& -\frac{1}{2^N} \sum_\mu \sum_{i<j} \xi^\mu_i \xi^\mu_j \sigma_i^z \sigma_j^z, 
\end{split}\\
\begin{split}
\label{eq:chop_h3}
   H_3^c =& -\frac{1}{2^N} \sum_\mu \sum_{i<j<l} \xi^\mu_i \xi^\mu_j \xi^\mu_l \sigma_i^z \sigma_j^z \sigma_l^z,  
\end{split}\\
\vdots& \nonumber
\end{align}
\end{subequations}
The second-order term $H_2^c$ in Eq.~(\ref{eq:chop_h2}) is precisely the Hamiltonian of the standard Hopfield model (up to an overall constant), which is known to possess a storage capacity that scales linearly with $N$ \cite{hopfield82, amit1986, amit1987}. The higher-order terms correspond to $p$-spin  modern Hopfield models, in which memories are encoded through $p$-body interactions \cite{hopfield2016}. These generalized models can achieve substantially larger capacities \cite{bovier2001, baldi1987, gardner1987}; in particular, a $p$-th order Hamiltonian can store up to $O(N^{p-1})$ stable patterns. Moreover, since the full projector Hamiltonian contains interactions of all orders, it naturally realizes the ``infinite"-order limit of dense associative memories. Upon neglecting subleading collision terms with repeated indices in the thermodynamic limit, the expansion of the projector Hamiltonian coincides with the expansion of an exponential interaction of the form $\sum_\mu \exp\left(\lambda \,\boldsymbol{\xi}^{\mu}\!\cdot\!\boldsymbol{\sigma} \right)$, revealing a close connection to the modern Hopfield network with exponential capacity \cite{mezard2024}. We have thereby established how the truncated quantum projector construction leads to the dense associative memory models.

% \textcolor{red}{KA: Can we make any statement about the p-> infty/exponential limit?}\ivar{didnt we conclude that the thing scales very similarly to exp(xi*sigma)? each term is basically $(\xi^\mu\cdot\sigma)^p/p!$, so they resum into $\exp(\xi^\mu\cdot\sigma)$}

\subsection{Quantum Dimer Associative Memory}
\label{section:quantum_dimer_model}
We now apply this construction to memories that are intrinsically quantum entangled. We consider the specific example where each memory is a dimer covering of $N$ spins, with each dimer $(ij)$ being a singlet pair,
\begin{equation}
\label{eq:dimer_covering}
    \ket{\psi^\mu_q} = \prod_{(ij) \in \mu} \frac{1}{\sqrt{2}} \left(\ket{\uparrow \downarrow}_{ij} - \ket{\downarrow \uparrow}_{ij}\right).
\end{equation}
Analogously to Eq.~(\ref{eq:chop_h_proj0}), we define the Hamiltonian as a sum of projectors onto the stored memories,
\begin{equation}
\label{eq:qhop_h_proj}
    H^q = -\sum_{\mu=1}^K \ket{\psi^\mu_q}\bra{\psi^\mu_q} = -\sum_\mu \prod_{(ij) \in \mu} \left(\frac{1}{4} - \mathbf{S}_{i} \!\cdot\! \mathbf{S}_{j}\right),
\end{equation}
where $\mathbf{S} = \frac{1}{2} \boldsymbol{\sigma}$ denotes the spin-$\frac{1}{2}$ operator, and $P_{ij} = \left(\frac{1}{4} - \mathbf{S}_{i} \!\cdot\! \mathbf{S}_{j}\right)$ is the projector onto a singlet formed by spins $i$ and $j$. Expanding the product of singlet projectors generates interactions of increasing order in $\boldsymbol{\sigma} \!\cdot\! \boldsymbol{\sigma}$. The first few nontrivial terms are
\begin{subequations}
\label{eq:qhop_h_expand}
\begin{align}
\begin{split}
\label{eq:qhop_h1}
H^q_1 =& + \frac{1}{2^N} \sum_\mu \sum_{(ij) \in \mu} \boldsymbol{\sigma}_{i} \!\cdot\! \boldsymbol{\sigma}_{j},
\end{split}\\
\begin{split}
\label{eq:qhop_h2}
H_2^q =& -\frac{1}{2^N} \sum_\mu \sum_{(ij)<(kl) \in \mu} \left(\boldsymbol{\sigma}_{i} \!\cdot\! \boldsymbol{\sigma}_{j}\right) \left(\boldsymbol{\sigma}_{k} \!\cdot\! \boldsymbol{\sigma}_{l}\right),
\end{split}\\
\begin{split}
\label{eq:qhop_h3}
H_3^q =& +\frac{1}{2^N} \sum_\mu \sum_{(ij)<(kl)<(pq) \in \mu} \left(\boldsymbol{\sigma}_{i} \!\cdot\! \boldsymbol{\sigma}_{j}\right) \left(\boldsymbol{\sigma}_{k} \!\cdot\! \boldsymbol{\sigma}_{l}\right)\left(\boldsymbol{\sigma}_{p} \!\cdot\! \boldsymbol{\sigma}_{q}\right),
\end{split}\\
\vdots& \nonumber
\end{align}
\end{subequations}
% \ivar{here in 6b and c the dimers should be sorted as well, right? $(ij)<(kl)$} 
The first-order term $H_1^q$ in Eq.~(\ref{eq:qhop_h1}) describes a collection of antiferromagnetic Heisenberg interactions on the bonds belonging to the stored memories. The second-order term $H_2^q$ in Eq.~(\ref{eq:qhop_h2}) introduces interactions between singlets and constitutes the quantum analogue of the pairwise Hopfield Hamiltonian. In the remainder of this paper, we focus primarily on $H_2^q$ and investigate its ability to store multiple entangled memory states (we briefly discuss the capacity of higher-order models at the end of Appendix \ref{section:two_spin_errors_capacity_appendix}).

\section{Local Energetic Stability Analysis}
\label{section:local_stability_analysis}
In this section, we examine the stability of individual stored memories with respect to various perturbations (unitary rotations). These perturbations  can mimic coupling to a low temperature bath -- if the energy of the memory perturbed by such rotation is lower than that of the original memory, then the memory will be destabilized by coupling to the bath. Just like in the classical Hopfield networks, we expect individual memories to become increasingly susceptible to destabilization as more memories are stored. The memory loading for which individual memories become unstable (probabilistically) defines the storage capacity. 

We consider perturbations of the form $e^{-i\epsilon \hat{O}_i}$, where $\hat{O}_i$ is a $k$-local Hermitian operator acting on at most $k$ spins. For $k=1$, these perturbations correspond to infinitesimal single-spin rotations (a magnetic-field bath). A memory state $\ket{\psi^\mu}$ is locally robust against a perturbation $\hat{O_i}$ if the perturbation leads to an increase in energy,
\begin{equation}
    \bra{\psi^\mu} e^{i\epsilon \hat{O}_i} H e^{-i\epsilon \hat{O}_i} \ket{\psi^\mu} - \bra{\psi^\mu} H \ket{\psi^\mu} > 0.
    \label{eq:energy_diff}
\end{equation}
Expanding Eq.~(\ref{eq:energy_diff}) to second order in $\epsilon$ yields the local stability conditions
\begin{subequations}
\label{eq:local_stability}
\begin{align}
\begin{split}
\bra{\psi^\mu} \left[\hat{O}_i, H\right] \ket{\psi^\mu} &= 0,
\label{eq:first_condition}
\end{split}\\
\begin{split}
    \bra{\psi^\mu} \left[\hat{O}_i, \left[H, \hat{O}_i\right]\right] \ket{\psi^\mu} &> 0.
    \label{eq:second_condition}
\end{split}
\end{align}
\end{subequations}
The first condition in Eq.~(\ref{eq:first_condition}) ensures that the memory state is a stationary point under the perturbation $\hat{O_i}$, while the second condition in Eq.~(\ref{eq:second_condition}) requires that this stationary point be a local minimum. If no local perturbations reduce the energy, the states are stable local minima of the energy landscape. Retrieval itself could be realized via coupling to the corresponding bath.

For the second-order Hamiltonian $H_2^q$, every memory is locally stable against arbitrary single-spin rotations. Qualitatively, a single-spin rotation breaks the singlet containing that spin and creates a triplet excitation. Since we are dealing with energy expectation values, we are essentially computing the energy difference
\[ 
\bra{t} \boldsymbol{\sigma}_i \cdot \boldsymbol{\sigma}_j \ket{t} - \bra{s} \boldsymbol{\sigma}_i \cdot \boldsymbol{\sigma}_j \ket{s},
\]
where $\ket{t}$ and $\ket{s}$ are triplet and singlet states of two spins. This difference is positive when $(ij)$ is the singlet bond affected by the rotation and vanishes when $(ij)$ is any other bond, independent of which stored memory a term in the Hamiltonian originates from. Since the memory under consideration always contains the affected singlet, it necessarily provides a positive contribution to the energy change. Thus, single-spin rotations cannot destabilize any memory (see Appendix~\ref{section:single_spin_rotations_appendix} for detailed explanation).
% \ivar{should we give a quick qualitative reason here as well? that single spin rotation messes up selected memory singlets, that is clear. But why other memories cannot produce noice that can mess things up?}

We further show that the only potentially unstable two-spin perturbations are local bond rearrangements that swap two spins between two different singlets, generated by operators of the form $\boldsymbol{\sigma}_{i} \! \cdot \! \boldsymbol{\sigma}_{j} = 2\mathrm{SWAP}_{ij} - I$. In the thermodynamic limit, the stability analysis for each bond rearrangement reduces to a combinatorial counting problem, in which one compares the numbers of stabilizing and destabilizing terms in $H_2^q$. A simple qualitative picture is as follows.  Suppose an error rearranges two singlets, from $(ik)(jl)$ to $(il)(jk)$, in a selected (``primary") memory. The stabilizing (energy-raising) terms are of the form $\left(\boldsymbol{\sigma}_{i} \! \cdot \!\boldsymbol{\sigma}_{k}\right) \left(\boldsymbol{\sigma}_{m} \!\cdot\! \boldsymbol{\sigma}_{n}\right)$, where $(ik)$ is one of the two original singlets affected by the bond rearrangement, and $(mn)$ is any unaffected singlet in the selected memory. Since there are $O(N)$ unaffected singlets, the selected memory contributes $O(N)$ stabilizing terms. Meanwhile, parts of the Hamiltonian that are associated with other (``secondary") memories can contribute destabilizing (energy-lowering) terms. These correspond to formation of pairs of singlets such as $\left(\boldsymbol{\sigma}_{j} \! \cdot \!\boldsymbol{\sigma}_{k}\right) \left(\boldsymbol{\sigma}_{m} \!\cdot\! \boldsymbol{\sigma}_{n}\right)$. A secondary memory must contain one of the two singlets newly created  by the swap, $(jk)$ or $(il)$, together with another singlet $(mn)$ shared with the primary memory (there are typically $O(1)$ singlets shared between two random memories), in order to contribute to lowering the energy. For a random memory, the probability of such an event therefore is $O(1/N)$. 
% for the newly created singlet to contribute to energy, that secondary memory should already contain a other singlets shared with the  primary memory.  There are typically $O(1)$ of those, each lowering energy by one unit.
There are also stabilizing terms arising from the secondary memories with the same probability. Together these terms ensure that the average contribution is zero. Since all secondary memories contribute independently, and with zero average, the total variance coming from all of them becomes $O(K/N)$ (we ignore the $2^{-N}$ scaling factor in Eqs. (\ref{eq:qhop_h1}- \ref{eq:qhop_h3}).
%In other words, as the system size increases, a random memory is increasingly unlikely to contain the specific bond pairing required to destabilize the selected memory. 
% \ivar{why is such a difference with classical, why N is in denominator? give a simple reason why increasing N reduces the noise}.
Balancing the energy increase due to the primary memory with the square root of the variance due to the secondary memories gives an estimate of capacity scaling as $K_\text{max}^q=O(N^3)$. A more careful analysis, which requires stability against all possible local bond rearrangements introduces a logarithmic correction and yields a capacity scaling as $K_\text{max}^q=O(N^3/\log N)$ (see Appendix~\ref{section:two_spin_errors_capacity_appendix} for detailed derivation). Analogous analysis yields capacity estimates scaling as $N^{2p_q-1}$ for all models $H^q_{p_q}$ with $p_q \geq 2$. $H_1^q$, on the other hand, does not possess an extensive capacity~\footnote{For a given pair of singlets $(ij)(kl)$ in a selected memory, a destabilizing contribution arises whenever another stored memory contains a rearranged pairing, e.g., $(ik)(jl)$, which occurs with probability $O(1/N^2)$ for a random perfect matching. Since there are $O(N^2)$ distinct pairs of singlets in a memory, the probability that at least one such destabilizing configuration exists is $O(1)$. Thus, $H_1^q$ does not possess an extensive storage capacity.}.

Presented capacity estimates parallel the ``signal-to-noise" analysis of the classical Hopfield model $H_{p_c}^c$, where the restoring signal is $O(N^{p_c-1})$, and the noise variance from other memories is $O(KN^{p_c-1})$, resulting in the familiar capacity scaling $K_\text{max}^c = O(N^{p_c-1})$ (we review the derivation for the classical Hopfield model in Appendix~\ref{section:capacity_classical_hopfield_appendix} for completeness).  Curiously, despite qualitative differences between quantum and classical models, the capacity of the quantum model scales as the same power of the system size as the classical model simply by identifying $2p_q$ with $p_c$. Whether this is just a coincidence or indicates a deeper connection between the classical and quantum models is not yet clear.
% \ivar{should we rename the capacity as $K_{max}$ and add another label q or c to distinguish quantum and classical?}

\begin{figure}
    \centering
    \subfloat[]{
    \includegraphics[width=0.47\linewidth]{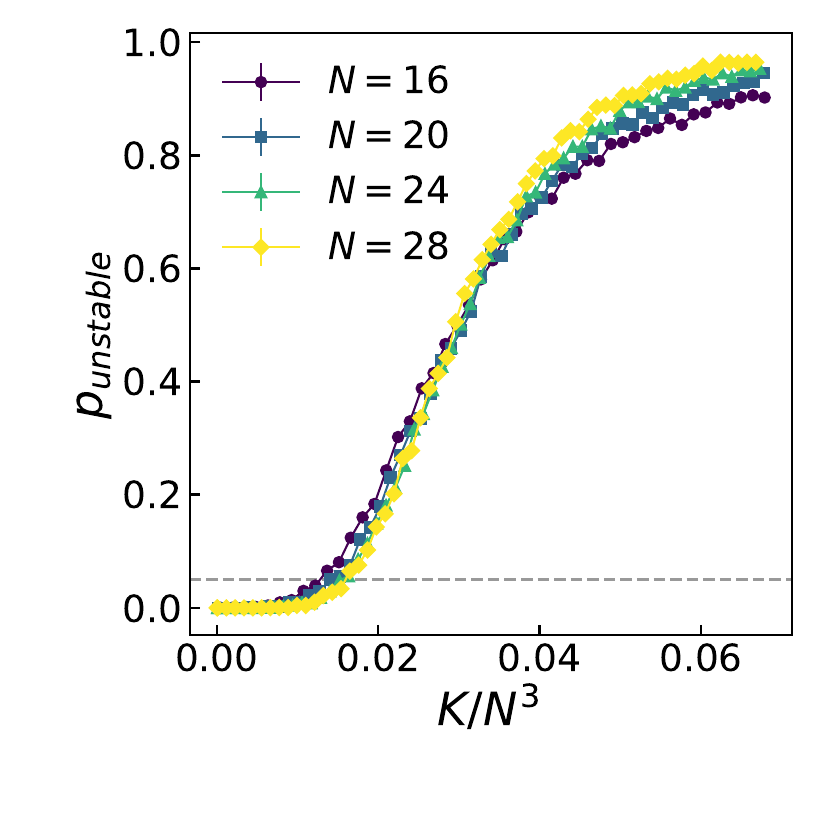}
    \label{fig:qhop_error_prob_scaling_h2}}
    \hfill
    \subfloat[]{
    \includegraphics[width=0.47\linewidth]{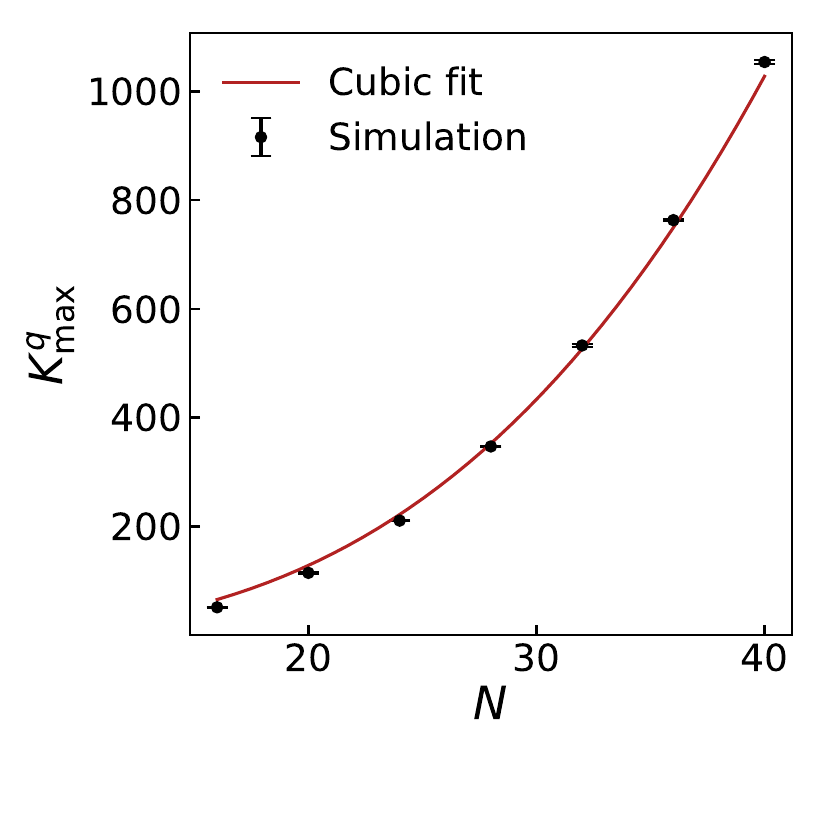}
    \label{fig:qhop_capacity_h2}}
    \caption{(a) The instability probability $p_{\mathrm{unstable}}(N,K)$ as a function of the rescaled memory load $K/N^3$ for different system sizes $N$. The dashed gray line indicates the threshold $p_0$ used to define the finite-size storage capacity $K_\text{max}^q$. (b) $K_\text{max}^q$, defined as the maximum number of stored memories for which the instability probability is below $p_0=5\%$, as a function of $N$. The data are well described by a cubic scaling, with a fitted coefficient of approximately $0.015$.}
    \label{fig:qhop_local_stability_h2}
\end{figure}

These statistical considerations can be verified by direct simulations. For each system size $N$, we generate a set of $K$ random memory states. We select one memory at random and test its stability against all possible bond rearrangements. If any such perturbation lowers the energy of the state, the memory is marked as unstable. Repeating this procedure over many independent realizations yields an estimate of the instability probability as a function of $N$ and $K$, 
\begin{equation}
    p_{\mathrm{unstable}}(N,K) = \Pr \! \left(\mathrm{instability}\right),
    \nonumber
\end{equation} 
Fig.~\ref{fig:qhop_error_prob_scaling_h2} 
% \textcolor{red}{KA: should hyperlink to the figure} 
shows $p_{\mathrm{unstable}}$ as a function of $K/N^3$ for different system sizes. The data collapse onto a single curve, providing numerical evidence for the predicted $N^3$ scaling of the storage capacity. For the system sizes accessible to our simulations, the logarithmic correction is too weak to be resolved numerically.

% This differs from the analytical lower bound by a factor of $\log N$, reflecting the fact that the union bound used in the analysis is not tight and overestimates the probability of instability because the errors are not independent. \textcolor{red}{KA: probably hard to claim we can tell logN dependence...} \ivar{i agree.}

In addition, we define the finite-size capacity as
\[
K_\text{max}^q(N)=\max\left\{K:\,p_{\mathrm{unstable}}(N,K) < p_0\right\},
\]
where $p_0$ is a fixed threshold. For our analysis, we choose $p_0=5\%$. The $N^3$ scaling is again confirmed by fitting the finite-size capacity as a function of the system size, as shown in Fig.~\ref{fig:qhop_capacity_h2}.

%\ivar{we shoudl think about reordering the follwoing sections. The Hybridization one is very convincing and nice, but may be we shoudl still go with Mattis first becasue it is the most natural}

\section{Quantum Hybridization of Memory States}
\label{section:quantum_hybridization}
The local stability analysis presented above is based on expectation values of the Hamiltonian and therefore provides a semiclassical criterion for memory stability, similar to the stability criteria in the classical Hopfield models. In a quantum system, however, distinct states can hybridize through off-diagonal matrix elements of the Hamiltonian, even when they have different energies. This issue obviously does not arise in the classical Hopfield model due to the absence of quantum tunneling. To quantify this genuinely quantum effect in our dimer model, we introduce a perturbative hybridization metric defined by the ratio between the coupling matrix element and the energy difference of two nearby dimer configurations.

Let us take a randomly chosen memory state $\ket{\psi^\mu_q}$. As shown in the last section, the elementary local perturbations are plaquette flips. We define the hybridization metric
\begin{equation}
    W = \max_{\widetilde{\mu}} \, \frac{ \left| \bra{\psi^\mu_q} H_2^q \ket{\phi_q^{\widetilde{\mu}}} \right| }{\left| E_\mu - E_{\widetilde{\mu}} \right|},
    \label{eq:hybrid_metric}
\end{equation}
where the maximum is taken over all dimer coverings $\widetilde{\mu}$ obtained from $\mu$ by a single plaquette flip. Since valence bond states are not orthogonal, we define
\begin{equation}
    \ket{\phi_q^{\widetilde{\mu}}} = \frac{\ket{\psi_q^{\widetilde{\mu}}} - \bra{\psi^\mu_q} \ket{\psi_q^{\widetilde{\mu}}} \ket{\psi^\mu_q}}{\sqrt{1 - \left|\bra{\psi^\mu_q} \ket{\psi_q^{\widetilde{\mu}}}\right|^2}}
    \nonumber
\end{equation}
so that $\ket{\phi_q^{\widetilde{\mu}}}$ is normalized and orthogonal to the original memory state. Here, $E_\mu$ and $E_{\widetilde{\mu}}$ denote the expectation values of the Hamiltonian in the states $\ket{\psi_q^\mu}$ and $\ket{\phi_q^{\widetilde{\mu}}}$, respectively.

To understand the motivation behind Eq.~(\ref{eq:hybrid_metric}), suppose we write the Hamiltonian as
\begin{equation}
    H_2^q = H_\mu + V,
    \nonumber
\end{equation}
where $H_\mu$ contains the terms associated with the selected memory $\mu$, and $V$ contains the contributions from all other memories. By construction, $\ket{\psi_q^\mu}$ is an eigenstate of $H_\mu$. Since $\ket{\phi_q^{\widetilde{\mu}}}$ has been orthogonalized against $\ket{\psi_q^\mu}$,
\[\bra{\psi_q^\mu}H_\mu\ket{\phi_q^{\widetilde{\mu}}}=0,\]
and
\[\bra{\psi_q^\mu} H_2^q \ket{\phi_q^{\widetilde{\mu}}}=\bra{\psi_q^\mu}V\ket{\phi_q^{\widetilde{\mu}}}.\]
Suppose we also treat $V$ perturbatively. First-order perturbation theory predicts that the amplitude for $\ket{\psi_q^\mu}$ to hybridize with $\ket{\phi_q^{\widetilde{\mu}}}$ is proportional to
\[
\frac{\bra{\psi_q^\mu}V\ket{\phi_q^{\widetilde{\mu}}}}
{E_\mu-E_{\widetilde{\mu}}}.
\]
This ratio is similar to the quantity that controls resonant hybridization in Anderson's localization criterion~\cite{Anderson1958}. In the present setting, the ``sites'' are points in the Hilbert space of dimer coverings rather than lattice sites in real space.
\begin{figure}
    \centering
    \subfloat[]{
    \includegraphics[width=0.47\linewidth]{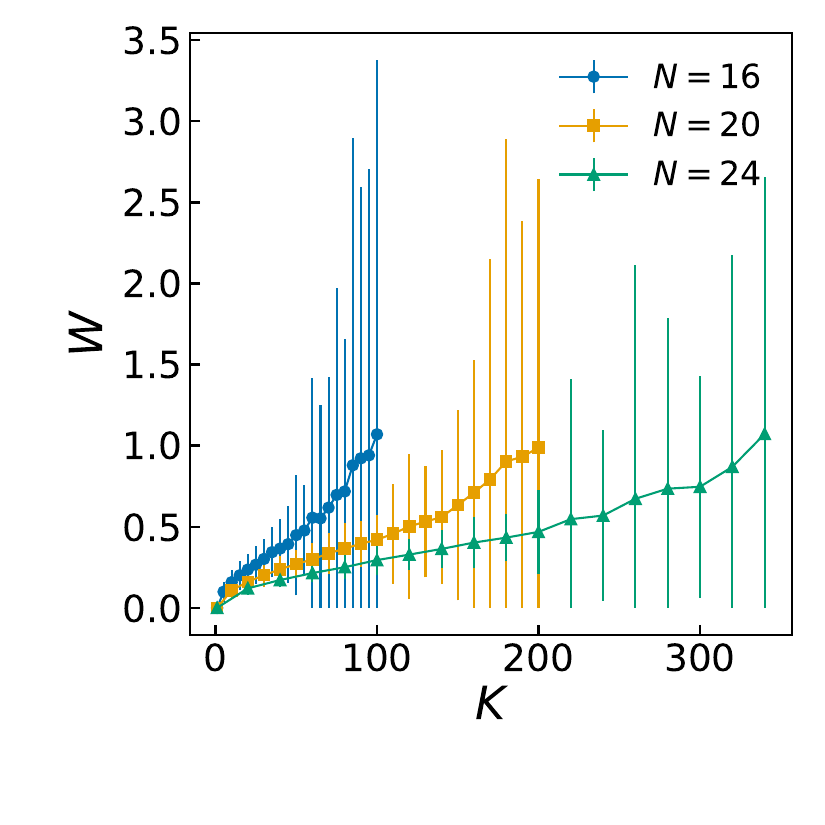}
    \label{fig:qhop_coupling_energy_ratio_errorbar_h2}}
    \hfill
    \subfloat[]{
    \includegraphics[width=0.47\linewidth]{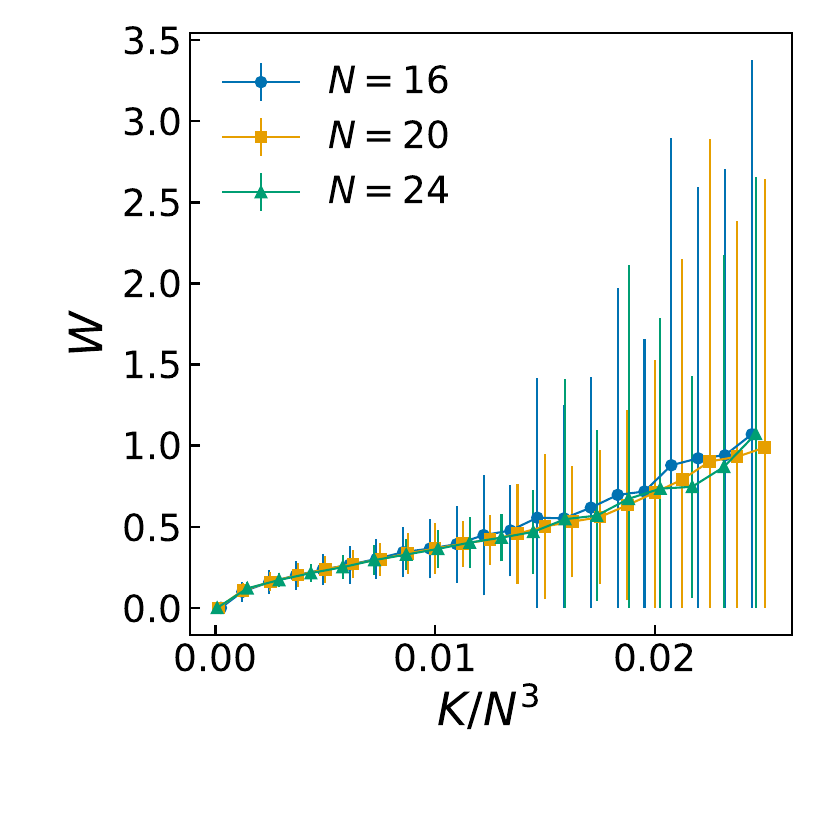}
    \label{fig:qhop_coupling_energy_ratio_scaled_errorbar_h2}}
    \caption{Hybridization metric $W$ defined in Eq.~\ref{eq:hybrid_metric} as a function of (a) the number of memories $K$ and (b) $K/N^3$. The error bars denote one standard deviation. The average collapses onto a single curve. The standard deviation increases sharply beyond the critical value $K_\text{max}^q/N^3 \approx 0.015$.}
    \label{fig:qhop_coupling_energy_ratio_h2}
\end{figure}

Fig.~\ref{fig:qhop_coupling_energy_ratio_h2} displays the hybridization metric $W$, averaged over many independent realizations of the Hamiltonian, for different system sizes and memory loadings. The data exhibit a finite-size collapse when plotted against $K/N^3$, consistent with the scaling predicted by the local stability analysis. Moreover, the standard deviation of $W$ grows rapidly beyond a critical value of $K_\text{max}^q \approx 0.015 N^3$, signaling the onset of resonant hybridization between a memory state and its nearby dimer coverings. The constant is also in agreement with that obtained from the fit in Fig.~\ref{fig:qhop_capacity_h2}.

\section{Eigenstate properties}
The previous two sections focused on the local stability of the memories, addressing when they become unstable to local perturbations and begin to hybridize with nearby dimer configurations. In this section, we instead characterize the memory content of the Hamiltonian eigenstates through their ``Mattis"  overlap with the stored memories and their entanglement structure.

We perform the analysis on a single eigenstate of the Hamiltonian $H_2^q$. We select the lowest-energy eigenstate within the $\boldsymbol{S}_{\text{tot}}^2=0$ sector, as the Hamiltonian preserves the total spin and the singlet sector is dynamically decoupled from the triplet sectors~\footnote{For the Hamiltonian $H_2^q$, states with aligned spins can become the ground state at sufficiently large memory loads $K$. Although each individual term in $H_2^q$ was designed to energetically favor singlet pairings, it assigns a negative energy to configurations in which pairs of spin pairs are aligned or anti-aligned. Then, as the number of stored memories increases, globally aligned spin states can satisfy more interaction terms simultaneously than any singlet configuration. Nevertheless, the singlet and triplet sectors are dynamically decoupled because the operators appearing in $H_2^q$ only reshuffle the singlet pairings and preserve the total spin.}. 

We include weak randomness in the coefficient of each memory in the Hamiltonian to avoid accidental degeneracies that could lead to resonant mixing between memories in finite systems. The form of the Hamiltonian we study is therefore
\begin{equation}
\begin{aligned}
\widetilde{H}_2^q
= - \sum_\mu \chi_\mu
\sum_{(ij)<(kl)\in\mu}
\left(\boldsymbol{\sigma}_{i}\!\cdot\!\boldsymbol{\sigma}_{j}\right)
\left(\boldsymbol{\sigma}_{k}\!\cdot\!\boldsymbol{\sigma}_{l}\right).
% \widetilde{H}_1^q+\widetilde{H}_2^q
% = - \sum_\mu \chi_\mu
% \Bigg[
% &-\sum_{(ij)\in\mu}
% \boldsymbol{\sigma}_{i}\!\cdot\!\boldsymbol{\sigma}_{j}
% \\
% &+
% \sum_{(ij)<(kl)\in\mu}
% \left(\boldsymbol{\sigma}_{i}\!\cdot\!\boldsymbol{\sigma}_{j}\right)
% \left(\boldsymbol{\sigma}_{k}\!\cdot\!\boldsymbol{\sigma}_{l}\right)
% \Bigg].
\end{aligned}
\label{eq:qhop_rand_h2}
\end{equation}
where $\chi_\mu \sim \mathcal{N} \left(1, 0.1 \right)$ is a random coefficient associated with memory $\mu$. 

% For numerical convenience, we consider the ground state of the Hamiltonian $H_1^q+H_2^q$ rather than $H_2^q$ alone. The terms in $H_1^q$ energetically penalize triplet sector which otherwise intervenes for larger $K$

% Adding $H_1^q$ does not change the scaling from the local stability analysis~\footnote{For a given pair of singlets $(ij)(kl)$ in a selected memory, a destabilizing contribution arises whenever another stored memory contains a rearranged pairing, e.g., $(ik)(jl)$, which occurs with probability $O(1/N^2)$ for a random perfect matching. Since there are $O(N^2)$ distinct pairs of singlets in a memory, the probability that at least one such destabilizing configuration exists is $O(1)$. Thus, $H_1^q$ by itself does not possess an extensive storage capacity.}.

% The local stability analysis of Sec.~\ref{section:local_stability_analysis} extends straightforwardly to this modified Hamiltonian. One can show that adding $H_1^q$ modifies the storage capacity only by a constant \ivar{constant factor? do we show that in the appendix?} and therefore does not change the $N^3$ scaling in the thermodynamic limit. 

\subsection{Mattis Overlap}
\label{section:mattis}
In the classical Hopfield network, the quality of the retrieved state is captured by the Mattis overlap \cite{amit1987}, which measures the fraction of spins aligned with the ideal stored pattern,
\begin{equation}
    m_\mu^c = \left\langle \frac{1}{N} \sum_i \xi_i^\mu \sigma_i^z \right\rangle.
\end{equation}
For a perfect retrieval state $\ket{\psi_c^\mu}$, one has $m_\mu^c=1$, while overlaps with  unrelated random patterns scale as $1/\sqrt{N}$,   vanishing in the thermodynamic limit.

By analogy, we define the singlet Mattis overlap of an eigenstate with a dimer memory $\mu$ as the average singlet weight on the bonds belonging to that memory,
\begin{equation}
    m_\mu^q = \left\langle \frac{2}{N} \sum_{(ij) \in \mu} P_{ij} \right\rangle,
    \label{eq:mattis_q}
\end{equation}
where $P_{ij} = \left(\frac{1}{4} - \mathbf{S}_{i} \!\cdot\! \mathbf{S}_{j}\right)$ projects onto a singlet bond $(ij)$ belonging to the memory $\mu$. This quantity equals unity for the memory state $\ket{\psi_q^\mu}$ and decreases as the state deviates from this dimer covering. For a covering state that is distinct from the selected memory $\mu$, $\langle P_{ij}\rangle$ equals $1$ if the bond $(ij)$ is shared with $\mu$ and $1/4$ otherwise. Since two random, independent dimer coverings share only $O(1)$ bonds on average, the contribution from the shared bonds vanishes as $N\to\infty$, giving
\[m_\mu^q \approx \frac{1}{4}\]
in the large-$N$ limit.

\begin{figure}
    \centering
    \includegraphics[width=0.9\linewidth]{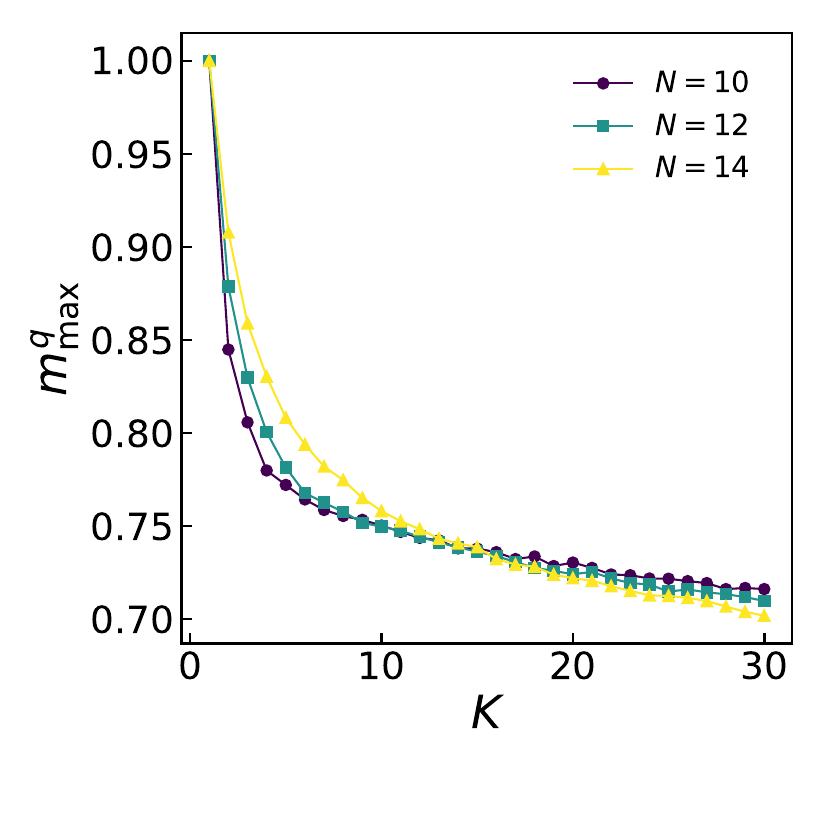}
    \caption{Maximum Mattis overlap between the lowest-energy eigenstate of $\widetilde{H}_2^q$ (Eq.~\ref{eq:qhop_rand_h2}) with $\boldsymbol{S}_{\text{tot}}^2=0$ and the stored memories as a function of memory load $K$ for different system sizes $N$. A transition appears to occur at some capacity $K_C (N)$ --- for $K < K_C (N)$, increasing $N$ enhances the overlap with a single memory, indicating the retrieval phase, while for $K > K_C (N)$, increasing $N$ decreases the overlap, indicating loss of memory. The capacity $K_C (N)$ also drifts to larger values with increasing $N$, as expected.}
    \label{fig:qhop_gs_mattis_overlap_max_h2}
\end{figure}

% For a state with no relation to any stored dimer covering, the two-spin reduced density matrix on a bond is essentially maximally mixed. Consequently,
% \[\langle P_{ij}\rangle \approx \frac{1}{4},\]
% so that
% \[m_\mu^q \approx \frac{1}{4}.\]

Fig.~\ref{fig:qhop_gs_mattis_overlap_max_h2} depicts the maximum singlet Mattis overlap
\[ m_\text{max}^q = \max_\mu \, m^q_\mu \]
between the lowest-energy eigenstate of $\widetilde{H}_2^q$ with $\boldsymbol{S}_{\text{tot}}^2=0$ and the stored memories for a range of memory loads $K$. For small $K$, the maximum overlap increases as the system size increases, meaning that the state becomes closer to one of the memories. As $K$ increases, the overlap decreases, and the state becomes less aligned with any individual stored dimer covering. The curves for different system sizes cross at distinct values of $K = K_C (N)$, suggesting a transition between a memory-dominated (related to retrieval) phase and a non-retrieval phase.

The $K_C (N) \propto N^3$ scaling cannot be resolved from this metric for the small system sizes accessible. In the non-retrieval phase, the maximum Mattis overlap is expected to decay as $1/4 + \mathcal{O} \left( \sqrt{\log K / N} \right)$. As we observe, however, the maximum remains appreciable $(\gtrsim 0.7)$, indicating that the eigenstate retains significant singlet characteristics of at least one stored memory even beyond the storage capacity. This is not unique to the quantum model, as seen from a similar analysis for the classical Hopfield model presented in Appendix~\ref{section:mattis_classical_hopfield}.

\subsection{Entanglement Structure}
\label{section:concurrence}
We employ two well-established entanglement measures, the one-tangle and the concurrence, to characterize how quantum correlations are distributed throughout the system in our dimer associative memory model. As in Sec.~\ref{section:mattis}, we focus on the lowest-energy state of the Hamiltonian $\widetilde{H}_2^q$ within the $\boldsymbol{S}_{\text{tot}}^2=0$ sector.

The one-tangle \cite{wootters2000} of spin $i$ is defined as 
\begin{equation}
    \tau_i = 4\,\mathrm{det}(\rho_i),
    \label{eq:one_tangle}
\end{equation}
where $\rho_i$ denotes the reduced density matrix of spin $i$. For a pure $N$-qubit state, the one-tangle quantifies the entanglement between a spin and the rest of the system. If a spin is completely unentangled, its reduced density matrix is pure, giving $\tau_i=0$. On the other hand, if it is maximally entangled with the other spins, its reduced density matrix is maximally mixed, $\rho_i=I/2$, and $\tau_i=1$. Thus, the one-tangle ranges from $0$ to $1$, with larger values indicating stronger entanglement between a single spin and the rest of the spins. 

The eigenstates within the $\boldsymbol{S}_{\text{tot}}^2=0$ sector are spanned by singlet dimer coverings which form an overcomplete basis. These states are clearly invariant under global SU(2) rotations of all spins. This implies that the reduced density matrix of every spin must be fully rotationally symmetric, which implies $\rho_i=I/2$. The one-tangle thus takes its maximal value, $\tau_i=1$, for every spin in these eigenstates. 

\begin{figure}
    \centering
    \subfloat[]{
    \includegraphics[width=0.47\linewidth]{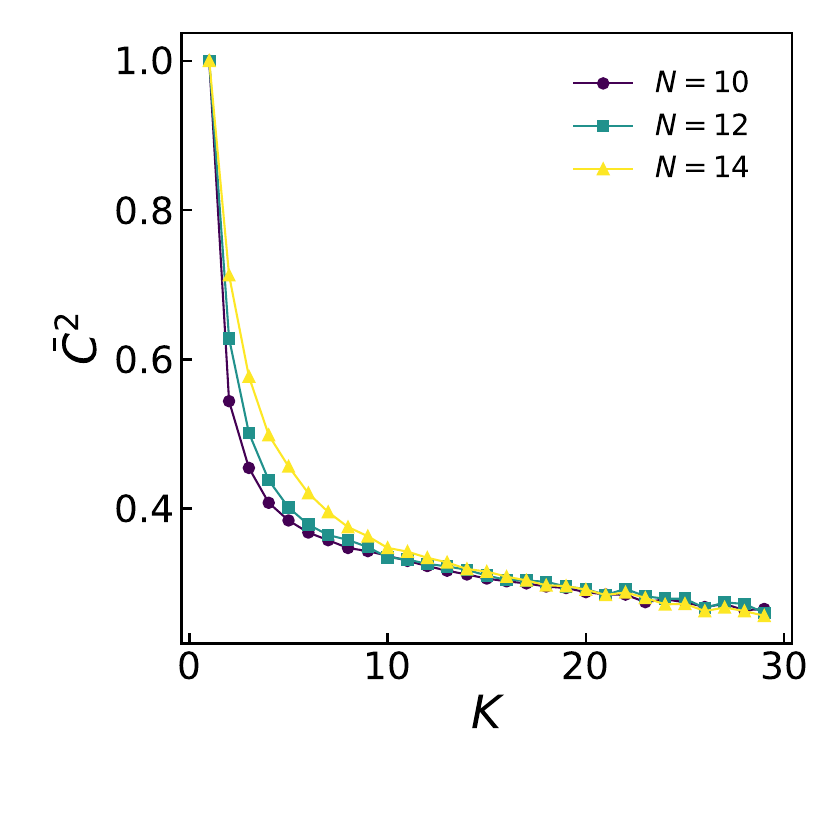}
    \label{fig:qhop_concur_sum_h2}}
    \hfill
    \subfloat[]{
    \includegraphics[width=0.47\linewidth]{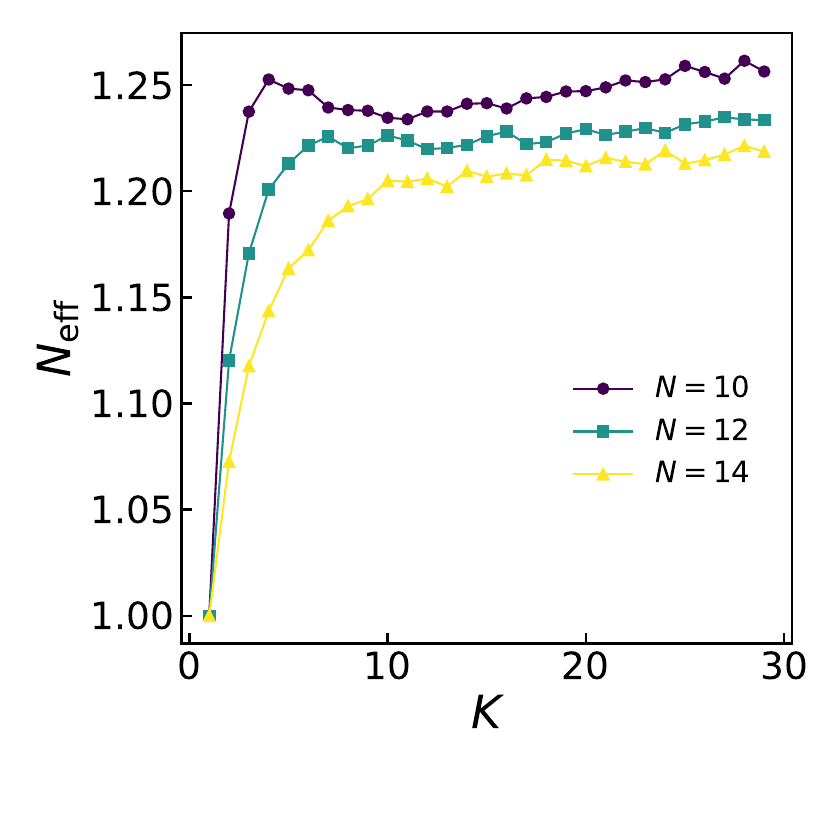}
    \label{fig:qhop_concur_ipr_h2}}
    \caption{(a) Average sum of squared concurrences $\Bar{C}^2$ (Eq.~\ref{eq:concur_sum}) and (b) effective number of entanglement partners $N_{\mathrm{eff}}$ (Eq.~\ref{eq:concur_ipr}) as a function of the memory load $K$ for different system sizes $N$. As $N$ increases, both quantities approach their values for a single dimer covering, $\Bar{C}^2=N_{\mathrm{eff}}=1$. Each spin has only $O(1)$ entangled partners, and in the thermodynamic limit the energy eigenstate approaches a single dimer covering in the retrieval phase (small $K$ region).}
    \label{fig:qhop_concur_h2}
\end{figure}

For every pair of spins $i \neq j$, we compute the two-spin reduced density matrix $\rho_{ij}$ and the corresponding concurrence \cite{wootters1997, wootters1998}
\begin{equation}
    C_{ij}=\max\!\left(0,\lambda_1-\lambda_2-\lambda_3-\lambda_4\right),
    \label{eq:concurrence}
\end{equation}
where $\lambda_1\geq\lambda_2\geq\lambda_3\geq\lambda_4$ are the eigenvalues of the Hermitian matrix
\[\sqrt{\sqrt{\rho_{ij}}\,\widetilde{\rho}_{ij}\,\sqrt{\rho_{ij}}},\]
with
\[ \widetilde{\rho}_{ij} = (\sigma_y\otimes\sigma_y)\,\rho_{ij}^{*}\,(\sigma_y\otimes\sigma_y). \]
The concurrence measures the bipartite entanglement between spins $i$ and $j$, taking values between $0$ for an unentangled pair and $1$ for a maximally entangled pair. For the rotationally invariant states considered here, the two-spin reduced density matrix is constrained to the form
\begin{equation}
\rho_{ij}=p_{ij}P_{ij}^s+\frac{1-p_{ij}}{3}P_{ij}^t,
\end{equation}
where $P_{ij}^s$ and $P_{ij}^t$ are the projectors onto the singlet and triplet sectors, respectively. The singlet weight is given by
\[
p_{ij}=\langle P_{ij}^s\rangle.
\]
The concurrence thus reduces to
\begin{equation}
C_{ij}=\max\!\left(0,2p_{ij}-1\right).
\end{equation}
The pairwise entanglement between spins $i$ and $j$ is completely determined by the expectation value of the singlet projector. A nonzero concurrence requires $p_{ij}>1/2$, providing a useful threshold for identifying an entangled bond. Although $p_{ij}$ is generally nonzero even when spins $i$ and $j$ do not form a singlet pair in a dimer covering or when the state is a superposition of coverings, values $p_{ij}\leq 1/2$ yield zero concurrence. Thus, the concurrence isolates pairs with sufficiently strong singlet character and provides a cleaner characterization of the pairwise entanglement structure than $p_{ij}$ alone.

The Coffman--Kundu--Wootters (CKW) inequality \cite{wootters2000,osborne2006} states that
\begin{equation}
    \sum_{j \neq i} C_{ij}^2 \leq \tau_i.
    \label{eq:ckw}
\end{equation}
The right-hand side represents the total entanglement between spin $i$ and the rest of the system. The left-hand side measures the portion of this entanglement that can be accounted for by pairwise correlations with the other spins. Any difference between the two arises from genuine multipartite entanglement that cannot be explained by pairwise contributions. For an ideal dimer covering, there is a unique spin $j$ such that $C_{ij}=1$ for each spin $i$, while all other concurrences involving spin $i$ vanish. Consequently,
\begin{equation}
\sum_{j\neq i}C_{ij}^2=\tau_i = 1,
\end{equation}
and the CKW inequality is saturated.

We define
\begin{equation}
    \widetilde{C}^2_{ij} = \frac{C_{ij}^2}{\sum_{k \neq i} C_{ik}^2},
\end{equation}
and the corresponding inverse participation ratio
\begin{equation}
    N_{\text{eff}}^{i} = \frac{1}{\sum_{j \neq i} \widetilde{C}_{ij}^4}.
\end{equation}
Averaging over all spins yields
\begin{equation}
    N_{\text{eff}} = \frac{1}{N} \sum_i N_{\text{eff}}^{i},
    \label{eq:concur_ipr}
\end{equation}
which quantifies the effective number of partners that a typical spin shares its pairwise entanglement with. For an ideal dimer covering, each spin has a single partner, giving $N_{\text{eff}}=1$. More generally, if the pairwise entanglement of a spin is distributed equally among $n$ partners, then $N_{\text{eff}}=n$. In Fig.~\ref{fig:qhop_concur_h2}, we plot both the average sum of squared concurrences
\begin{equation}
    \Bar{C}^2 = \frac{1}{N} \sum_i \left( \sum_{j \neq i} C_{ij}^2 \right),
    \label{eq:concur_sum}
\end{equation}
and the inverse participation ratio $N_{\text{eff}}$. Both quantities approach their ideal dimer-covering values as the system size increases in the retrieval phase (small $K$), indicating that the pairwise entanglement becomes increasingly concentrated into a single partner for each spin. This behavior suggests that the eigenstate approaches the structure of a single dimer covering in the thermodynamic limit.

\section{Discussion}
% \textcolor{red}{KA: I think first paragraph should state all results of previous sections as a summary and comparison with standard classical Hopfield model. Then you can go into how it can be used to store relational structures. }
In this paper, we introduced an associative memory model that stores entangled quantum many-body states. We investigated its storage capacity from several complementary perspectives, including local energetic stability and quantum hybridization. Our analysis shows that the storage capacity scales as $N^3$, substantially exceeding the linear scaling expected by naive analogy with the standard classical Hopfield model. We also showed that, in the retrieval phase, the Hamiltonian eigenstates are increasingly dominated by individual dimer-covering memories.

An important feature of our construction is that the memories encode perfect matching information through entangled valence-bond states. Such information cannot be reduced to classical local spin orientations since a singlet is fundamentally not reducible to a product state of constituent spins.
Quantum entanglement therefore enables associative memories to naturally store relational (graph) structures, not easily representable by spin product states.

Several questions remain open. First, the current model stores only a very simple kind of quantum entangled states with a valence bond structure. It would be interesting to apply our framework to accommodate other classes of entangled memories, such as random stabilizer graph states \cite{hein2006}. Second, our analysis focused essentially on static properties of closed quantum systems. Including  dynamics and dissipation would be necessary to study how such networks can be used in practice for storage and retrieval of entangled memories, analogous to the dynamical mechanisms employed in classical Hopfield networks \cite{rotondo2018, torres2024, kimura2025}. We leave these question for future investigations.

%\textcolor{red}{KA: this is really speculative. I would remove. Or we think a bit more and then expand and write with confidence.} \ivar{Probably agree. We should end on clear point. What about making it about applications -- where such quantum matching problems may occur, logistics, etc? If we mention error decoding in that context, that shoudl be fine. But if we make it all about error correction, then the standard is much higher for explaining what we have in mind} Finally, an interesting direction is the application of quantum associative memories to quantum error correction. In a QEC setting, the stored memories could represent error configurations associated with different syndromes. Given an observed syndrome and an initial guess of the error, the associative memory could retrieve the most likely error pattern.

\begin{acknowledgments}
We gratefully acknowledge support by the US Department of Energy, Office of Science, Basic Energy Sciences, Materials Sciences and Engineering Division. The authors used ChatGPT for language editing and improving the clarity of the manuscript. All scientific content, analysis, and conclusions are the authors' own.
\end{acknowledgments}

\appendix

\section{Local Energetic Stability and Storage Capacity of Quantum Dimer Associative Memory}
\label{section:local_stability_quantum_dimer_appendix}
In this appendix, we present a detailed analysis of the local stability conditions and derive the capacity scaling for the quantum dimer associative memory model outlined in Sec.~\ref{section:local_stability_analysis}. We consider the conditions
\begin{subequations}
\label{eq:local_stability_append}
\begin{align}
\begin{split}
\bra{\psi^\mu_q} \left[\hat{O}_i, H_2^q\right] \ket{\psi^\mu_q} &= 0,
\label{eq:first_condition_append}
\end{split}\\
\begin{split}
    \bra{\psi^\mu_q} \left[\hat{O}_i, \left[H_2^q, \hat{O}_i\right]\right] \ket{\psi^\mu_q} &> 0.
    \label{eq:second_condition_append}
\end{split}
\end{align}
\end{subequations}
for perturbations generated by local operators $\hat{O_i}$, applied to the Hamiltonian (ignoring the overall constant)
\begin{equation}
    \label{eq:qhop_h2_append}
    H_2^q = - \sum_\mu \sum_{(ij)<(kl) \in \mu} \left(\boldsymbol{\sigma}_{i} \!\cdot\! \boldsymbol{\sigma}_{j}\right) \left(\boldsymbol{\sigma}_{k} \!\cdot\! \boldsymbol{\sigma}_{l}\right),
\end{equation}
where each memory $\mu$ is a perfect pair matching of $N$ spins, represented by the valence bond state
\begin{equation}
    \label{eq:dimer_covering_append}
    \ket{\psi^\mu_q} = \prod_{(ij) \in \mu} \frac{1}{\sqrt{2}} \left(\ket{\uparrow \downarrow}_{ij} - \ket{\downarrow \uparrow}_{ij}\right).
\end{equation}

\subsection{Stability under Single-Spin Rotations}
\label{section:single_spin_rotations_appendix}
We first consider the stability of memory states under single-spin perturbations generated by local rotations $\sigma^\alpha_i$, where $\alpha \in \{x,y,z\}$ denotes the different Pauli operators. The first-order condition is
\begin{equation}
\bra{\psi^\mu_q} \left[\sigma_i^\alpha, H_2^q\right] \ket{\psi^\mu_q} = 0,
\qquad \forall \alpha, i.
\label{eq:first_condition_single_spin}
\end{equation}
The second-order condition is
\begin{equation}
    M_i \succ 0 \qquad \forall i,
    \label{eq:second_condition_single_spin}
\end{equation}
where $M_i$ is the $3 \times 3$ matrix for each spin $i$ with elements
\begin{equation}
    M_i^{\alpha\beta} = \bra{\psi^\mu_q} \left[\sigma_i^\alpha, \left[H_2^q, \sigma_i^\beta\right]\right] \ket{\psi^\mu_q}.
    \label{eq:second_condition_single_spin_ele}
\end{equation}
Local stability requires $M_i$ to be positive definite.

Note that for any dimer covering state $\ket{\psi^\mu_q}$, $\langle \sigma_i^\alpha \rangle = 0$. $\langle \sigma_i^\alpha \sigma_j^\beta \rangle = -\delta^{\alpha\beta}$ whenever spins $i$ and $j$ form a singlet pair in $\ket{\psi^\mu_q}$; otherwise the expectation value vanishes. More generally, the expectation value of any operator string containing an odd number of sites vanishes. For an even number of sites, the expectation value is nonzero only if the Pauli operators occur in singlet pairs with matching components on the two spins of each singlet.

The left hand side of Eq.~(\ref{eq:first_condition_single_spin}) is
\begin{eqnarray}
    && \bra{\psi^\mu_q} \left[\sigma_i^\alpha, H_2^q\right] \ket{\psi^\mu_q} \nonumber \\
    % && \qquad \quad = -\sum_\nu \sum_{(mn)<(kl) \in \nu} \left\langle \left[\sigma_i^\alpha, \left(\boldsymbol{\sigma}_{m} \!\cdot\! \boldsymbol{\sigma}_{n}\right) \left(\boldsymbol{\sigma}_{k} \!\cdot\! \boldsymbol{\sigma}_{l}\right)\right] \right\rangle \nonumber \\
    && \quad = -\sum_\nu \sum_{(mn)<(kl) \in \nu} \left\langle \left[\sigma_i^\alpha, \sigma_m^\beta \sigma_n^\beta \sigma_k^\gamma \sigma_l^\gamma \right] \right\rangle.
    \label{eq:first_condition_single_spin_h2}
\end{eqnarray}
Every term in the sum vanishes. If $i \notin \{m,n,k,l \}$, the operators commute and the commutator is identically zero. If $i \in \{m,n,k,l \}$, the commutator generates operator strings that either contain an odd number of sites or contain non-matching Pauli components within at least one singlet pair. The expectation value of every such term vanishes. Therefore, the first stability condition is always satisfied.

For the second condition, Eq.~(\ref{eq:second_condition_single_spin_ele}) reads
\begin{equation}
    M_i^{\alpha\beta}
    = -\sum_\nu \sum_{(mn)<(kl) \in \nu} \left\langle \left[\sigma_i^\alpha, \left[\sigma_m^\rho \sigma_n^\rho \sigma_k^\gamma \sigma_l^\gamma, \sigma_i^\beta\right]\right] \right\rangle.
\label{eq:second_condition_single_spin_h2}
\end{equation}
Since both $H_2^q$ and $\ket{\psi^\mu_q}$ are invariant under global SU(2) rotations, $M_i$ must be proportional to $\delta^{\alpha\beta}$. It is therefore sufficient to consider the diagonal case $\alpha = \beta$ with $i \in \{m, n, k, l\}$. In this case, the double commutator produces non-vanishing contributions only when $\{m, n, k, l\}$ belong to two singlets in memory $\mu$. For such terms, the correlator factorizes over singlets, $\left\langle \sigma \sigma \sigma \sigma \right\rangle = \left\langle \sigma \sigma \right\rangle \left\langle \sigma \sigma \right\rangle = (-1)(-1) = 1$. By construction, the Hamiltonian contains terms associated with the memory $\mu$, ensuring that at least one such positive contribution is present. Contributions from the other memories are also non-negative and can only increase the energy difference. Hence, $M_i^{\alpha\alpha}$ is always positive, and $M_i$ is positive definite. The second condition is also always satisfied.

What we have shown is that any single-spin rotation necessarily breaks a singlet pair and produce a triplet excitation orthogonal to the memory manifold. These local excitations always increase the energy relative to the original dimer covering state, so the memory states are stable local minima of the energy landscape with respect to arbitrary single-spin perturbations.

\subsection{Stability under Two-Spin Correlated Deformations and Storage Capacity}
\label{section:two_spin_errors_capacity_appendix}
We now consider errors acting on pairs of spins, generated by local operators $\sigma_i^\alpha \sigma_j^\beta$. These perturbations probe whether memory stability extends beyond single-spin excitations to correlated  deformations of the valence-bond structure.

The stability conditions follow from evaluating
\begin{eqnarray}
    &&\bra{\psi^\mu_q} \left[\sigma_i^\alpha \sigma_j^\beta, H_2^q\right] \ket{\psi^\mu_q} = \nonumber \\
    &&\quad -\sum_\nu \sum_{(mn)<(kl) \in \nu} \left\langle \left[\sigma_i^\alpha \sigma_j^\beta, \sigma_m^\rho \sigma_n^\rho \sigma_k^\gamma \sigma_l^\gamma \right] \right\rangle,
    \label{eq:first_condition_two_spin_h2}
\end{eqnarray}
and
\begin{align}
\begin{split}
M_{ij}^{\alpha\beta\gamma\delta}=\bra{\psi^\mu_q} \left[\sigma_i^\alpha \sigma_j^\beta, \left[H_2^q, \sigma_i^\gamma \sigma_j^\delta \right]\right] \ket{\psi^\mu_q} \nonumber \end{split} \\
\begin{split}
    \qquad = -\sum_\nu \sum_{(mn)<(kl) \in \nu} \left\langle \left[\sigma_i^\alpha \sigma_j^\beta, \left[\sigma_m^\rho \sigma_n^\rho \sigma_k^\eta \sigma_l^\eta, \sigma_i^\gamma \sigma_j^\delta \right]\right] \right\rangle.
\end{split}
\label{eq:second_condition_two_spin_h2}
\end{align}

Eq.~(\ref{eq:first_condition_two_spin_h2}) evaluates to zero. Either the operators commute and the commutator is identically zero, or the commutator generates operator strings that, when evaluated in the dimer state $\ket{\psi^\mu_q}$, contain at least one singlet on which an unpaired
(or mismatched) Pauli operator acts. Such contributions vanish, and we conclude that the first stability condition is always satisfied.

The second condition requires that the $9 \times 9$ matrix $M_{ij}$, with elements defined in Eq.~(\ref{eq:second_condition_two_spin_h2}), be positive definite. We first consider the case in which spins $i$ and $j$ form a singlet pair in the memory state $\mu$. The expectation value is non-vanishing only when ($\alpha = \gamma, \beta = \delta, \alpha \neq \beta$) or ($\alpha = \delta, \beta = \gamma, \alpha \neq \beta$). Taking the operators $\sigma^x_i\sigma^z_j$ and $\sigma^z_i\sigma^x_j$ as an example, the corresponding block in $M_{ij}$ is a $2 \times 2$ matrix
\begin{equation}
\left(
\begin{array}{cc}
    a & -a \\
    -a & a
\end{array}
\right), \nonumber
\end{equation}
where $a$ is a positive constant. This matrix has eigenvalues $0$ and $2a$. The zero mode corresponds to the eigen-operator $\sigma^x_i\sigma^z_j + \sigma^z_i\sigma^x_j$, which annihilates the singlet,
\begin{equation}
    (\sigma^x_i\sigma^z_j + \sigma^z_i\sigma^x_j)\ket{\text{singlet}}_{ij} = 0. \nonumber
\end{equation}
It does not indicate a physical instability, but rather reflects a redundant perturbation that acts trivially within the singlet subspace.

Next, we consider the case in which spins $i$ and $j$ do not form a singlet pair in $\mu$. The matrix $M_{ij}$ in this case is block diagonal. Let $k$ and $l$ denote the singlet partners of $i$ and $j$, respectively, in $\mu$, and let $(m,n)$ denote any other singlet pair in the same memory. One block consists of the operators $\sigma_i^x \sigma_j^x$, $\sigma_i^y \sigma_j^y$ and $\sigma_i^z \sigma_j^z$, and is given by
\begin{equation}
\left(
\begin{array}{ccc}
    24a+16c+8f & -12b-8d-4g & -12b-8d-4g \\
    -12b-8d-4g & 24a+16c+8f & -12b-8d-4g \\
    -12b-8d-4g & -12b-8d-4g & 24a+16c+8f
\end{array}
\right),
\label{eq:stability_matrix}
\end{equation}
where
\begin{enumerate}
    \item $a$ is the number of bond pairs of the form $(ik)(mn)$ or $(jl)(mn)$ over all stored memories,
    \item $b$ is the number of bond pairs of the form $(il)(mn)$ or $(jk)(mn)$ over all stored memories,
    \item $c$ is the number of bond pairs of the form $(ik)(jl)$ over all stored memories,
    \item $d$ is the number of bond pairs of the form$(il)(jk)$ over all stored memories,
    \item $f$ is the number of bond pairs of the form $(im)(kn)$ or $(jm)(ln)$ over all stored memories,
    \item $g$ is the number of bond pairs of the form $(im)(ln)$ or $(jm)(kn)$ over all stored memories.
\end{enumerate}
Diagonalizing this matrix reveals at most one potentially unstable direction, corresponding to the operator $\sigma_i^x \sigma_j^x + \sigma_i^y \sigma_j^y + \sigma_i^z \sigma_j^z$. By the identity $\boldsymbol{\sigma}_i \!\cdot\! \boldsymbol{\sigma}_j = 2\mathrm{SWAP}_{ij} - I$, this operator exchanges the quantum states of spins $i$ and $j$ belonging to different singlets, thereby rearranging the singlet bonds. This result simplifies the stability analysis by identifying bond rearrangements as the only relevant local two-spin instability.

The remaining blocks of $M_{ij}$ are $2 \times 2$ matrices, corresponding to pairs of operators such as $\sigma^x_i\sigma^z_j$ and $\sigma^z_i\sigma^x_j$. Their matrix elements can be expressed in terms of combinatorial counts outlined above. These blocks do not introduce any additional instability: whenever one of their eigenvalues becomes negative, the eigenvalue associated with the operator $\boldsymbol{\sigma}_i \!\cdot\! \boldsymbol{\sigma}_j$ is already negative. It is sufficient to analyze only the bond rearrangements.

% \subsection{Capacity Scaling}
% \ivar{Capacity and stability are not two different things. Feels strange to have this as an appendix separate form A2} \textcolor{blue}{Makes sense, we can merge them.} \ivar{lets do that}
% \label{section:capacity_quantum_dimer_appendix}
Consider a local bond rearrangement generated by the operator $\boldsymbol{\sigma}_i \!\cdot\!\boldsymbol{\sigma}_j$, which transforms the bond configuration
$(ik)(jl)$ into $(il)(jk)$. To estimate the storage capacity, we examine how the entries of the matrix in Eq.~(\ref{eq:stability_matrix}) scale with the system size. The dominant contribution to stability comes from bond pairs of the form $(ik)(mn)$ or $(jl)(mn)$. By construction, the Hamiltonian contains $O(N)$ such terms associated with the selected memory $\mu$. In contrast, bond pairs of the form $(il)(mn)$ or $(jk)(mn)$ from the other memories contribute de-stabilizing terms. The remaining classes of terms receive only $O(1)$ contributions from each individual memory. Since they lack the extensive $O(N)$ contributions, they do not determine the leading scaling of the capacity. Consequently, the $a$- and $b$-terms dominate the matrix in the thermodynamic limit, allowing it to be approximated by
\begin{equation}
\left(
\begin{array}{ccc}
    24a & -12b & -12b \\
    -12b & 24a & -12b \\
    -12b & -12b & 24a
\end{array}
\right). \nonumber
% \label{eq:stability_matrix_approx}
\end{equation}
The eigenvalue corresponding to the potentially unstable direction is simply $24(a-b)$. Checking the local stability of the quantum dimer memories against local plaquette flips reduces to a simple combinatorial problem: counting the number of stabilizing contributions, represented by $a$, and the number of competing destabilizing contributions, represented by $b$, arising from all stored memories.

For the selected memory $\mu$, there are $t = N-4$
stabilizing bond pairs of the form $(ik)(mn)$ or $(jl)(mn)$, where $(mn)$ is any other singlet pair in $\mu$. Let $S$ and $D$ denote the total numbers of stabilizing and destabilizing terms from all other stored memories. The bond rearrangement becomes locally unstable only if the destabilizing contributions outweigh the stabilizing ones, namely, $D-S \geq t$. To estimate the probability of this event, we define the random variables $X_\nu$, $S_\nu$, and $D_\nu$ for each random memory $\nu$, where $X_\nu = D_\nu - S_\nu$, and $X = \sum_{\nu \neq \mu} X_\nu = D - S$. $D_\nu = \sum_{(mn) \in \mu} \mathbf{1}_{(il)(mn) \in \nu} + \mathbf{1}_{(jk)(mn) \in \nu}$, where $\mathbf{1}_{(il)(mn) \in \nu}$ is the indicator variable taking the value $1$ if the bond pairs $(il)$ and $(mn)$ both belong to the memory $\nu$, and $0$ otherwise.

We first compute the expectation value of $D_\nu$:
\begin{align}
\mathbb{E}[D_\nu]
&= \sum_{(mn)\in\mu}
\mathbb{E}[\mathbf{1}_{(il)(mn)\in\nu}]
+\mathbb{E}[\mathbf{1}_{(jk)(mn)\in\nu}]
\nonumber\\
&=2\left(\frac N2-2\right)\frac1{(N-1)(N-3)} \nonumber\\
&=\frac{N-4}{(N-1)(N-3)}
\sim\frac1N .
\label{eq:expect_val_d_nu}
\end{align}
Similarly, the expectation value of $D_\nu^2$ is
\begin{widetext}
\begin{align}
\begin{split}
    \mathbb{E}\!\left[ D_\nu^2 \right] &= \sum_{(mn) \in \mu} \mathbb{E}\!\left[ \mathbf{1}_{(il)(mn) \in \nu} \right] + \mathbb{E}\!\left[ \mathbf{1}_{(jk)(mn) \in \nu} \right] 
    + \sum_{(mn) \neq (pq) \in \mu} \mathbb{E}\! \left[ \mathbf{1}_{(il)(mn) \in \nu} \, \mathbf{1}_{(il)(pq) \in \nu} \right]
\end{split} \nonumber \\
\begin{split}
    & \qquad \quad + \sum_{(mn) \neq (pq) \in \mu} \mathbb{E}\! \left[ \mathbf{1}_{(jk)(mn) \in \nu} \, \mathbf{1}_{(jk)(pq) \in \nu} \right]
    + 2 \sum_{(mn),(pq) \in \mu} \mathbb{E} \!\left[ \mathbf{1}_{(il)(mn) \in \nu} \, \mathbf{1}_{(jk)(pq) \in \nu} \right]
\end{split} \nonumber \\
\begin{split}
    &=  2 \left( \frac{N}{2} - 2 \right) \frac{1}{(N-1)(N-3)} + 2 \left( \frac{N}{2} - 2 \right) \left( \frac{N}{2} - 2 - 1 \right) \frac{1}{(N-1)(N-3)(N-5)}
\end{split} \nonumber \\
\begin{split}
    &\qquad \quad + 2 \left( \frac{N}{2} - 2 \right) \frac{1}{(N-1)(N-3)(N-5)} + 2\left( \frac{N}{2} - 2 \right) \left( \frac{N}{2} - 2 - 1 \right) \frac{1}{(N-1)(N-3)(N-5)(N-7)}
\end{split} \nonumber \\
\begin{split}
    & \sim \frac{3}{2N}.
\end{split}
\label{eq:var_d_nu}
\end{align}
\end{widetext}
The same expectation values hold for $S_\nu$. Note that $D_\nu$ and $S_\nu$ are mutually exclusive: a random memory contributes either stabilizing or destabilizing terms, but not both. Therefore, $\mathbb{E}\! \left[ D_\nu S_\nu \right] = 0$. We obtain
\begin{equation}
    \mathbb{E}\! \left[ X_\nu \right]= \mathbb{E}\! \left[ D_\nu \right] - \mathbb{E}\! \left[ S_\nu \right] = 0,
    \label{eq:expect_val_x_nu}
\end{equation}
and 
\begin{equation}
    \mathbb{E}\! \left[ X_\nu^2 \right] = \mathbb{E}\! \left[ D_\nu^2 \right] + \mathbb{E}\! \left[ S_\nu^2 \right] \sim \frac{3}{N}.
\label{eq:var_x_nu}
\end{equation}

Although the maximum possible value of $X_\nu$ is $O(N)$, such large deviations are exponentially unlikely for a random memory. It is convenient to separate the typical and rare contributions by defining
\begin{equation}
    \widetilde{X}_\nu = X_\nu \, \mathbf{1}_{|X_\nu| < r}, \qquad R_\nu = X_\nu \, \mathbf{1}_{|X_\nu| \geq r}, \nonumber
\end{equation}
where $r$ is a threshold satisfying $r = o(N)$. The probability of instability can be bounded by
\begin{widetext}
\begin{align}
\begin{split}
    \Pr \! \left( \sum_{\nu \neq \mu}X_\nu \geq t \right) &\leq \Pr \! \left( \sum_{\nu \neq \mu}\widetilde{X}_\nu \geq \frac{t}{2} \right) + \Pr \! \left( \sum_{\nu \neq \mu} R_\nu \geq \frac{t}{2} \right)
\end{split} \nonumber \\
\begin{split}
    &\leq \Pr \! \left( \sum_{\nu \neq \mu}\widetilde{X}_\nu \geq \frac{t}{2} \right) + \Pr\!  \left( \exists \, \nu \neq\mu: R_\nu \neq 0 \right) \\
    &= \Pr\!\left(\sum_{\nu\neq\mu}\widetilde{X}_\nu \geq \frac{t}{2}\right) + \Pr \!\left(\exists\,\nu\neq\mu:\,|X_\nu| \geq r\right).
\end{split}
\label{eq:prob_x_nu}
\end{align}
\end{widetext}

We analyze the two terms in Eq.~(\ref{eq:prob_x_nu}) separately. By the union bound,
\begin{equation}
    \Pr \!\left(\exists\,\nu\neq\mu:\,|X_\nu| \geq r\right) \leq (K-1) \Pr\! \left(|X_\nu| \geq r\right).
\label{eq:prob_r_nu_union_bound}
\end{equation}
The event $|X_\nu| \geq r$ occurs only if either $D_\nu \geq r$ or $S_\nu \geq r$. These two events have the same probability, giving an overall factor of $2$. Let $A_\nu$ denote the event that the bond $(ik)$ belongs to the random memory $\nu$, and $B_\nu$ the event that $(jl)$ belongs to $\nu$. Let $Y_\nu$ denote the number of additional bond pairs $(mn) \in \mu$ that also belong to $\nu$. Then,
\begin{widetext}
\begin{align}
\begin{split}
    \Pr \! \left(|X_\nu| \geq r\right) &= 2 * \left\{ \Pr\! \left( A_\nu \oplus B_\nu \right) \, \Pr\! \left( Y_\nu \geq r \, \middle| \, A_\nu \oplus B_\nu \right) + \Pr\! \left( A_\nu \cap B_\nu \right)  \Pr\! \left( Y_\nu \geq \frac{r}{2} \, \middle| \, A_\nu \cap B_\nu \right)\right\}.
\end{split} \nonumber \\
\begin{split}
    &= 2 * \left\{ \frac{2(N-4)}{(N-1)(N-3)} \, \Pr\! \left( Y_\nu \geq r \, \middle| \, A_\nu \oplus B_\nu \right) + \frac{1}{(N-1)(N-3)} \, \Pr\! \left( Y_\nu \geq \frac{r}{2} \, \middle| \, A_\nu \cap B_\nu \right) \right\}.
\end{split}
\label{eq:prob_r_nu_single}
\end{align}
\end{widetext}

We first compute the probability that a random perfect matching $M$ on $2m$ vertices contains exactly $y$ prescribed edges. Let $d_1, d_2, \ldots, d_L$ denote the prescribed edges. On $2m$ vertices, there are $(2m-1)!!$ perfect matchings. If we choose a subset of $y$ prescribed edges to appear in $M$, $G \subset \{d_1, d_2, \ldots, d_L \}$ with $|G|=y$, the remaining $2m-2y$ vertices may be matched arbitrarily. Hence, 
\begin{equation}
    \Pr \! \left( G \subset M \right) = \frac{(2m-2y-1)!!}{(2 m-1)!!}. 
    \nonumber
\end{equation}
To get the probability of having \emph{exactly} the edges in $G$, we must additionally require that none of the remaining $L-y$ prescribed edges appear. Applying the inclusion--exclusion principle and summing over all $\binom{L}{y}$ choices of $G$ yields
\begin{equation}
\begin{split}
\Pr\!\left(\text{exactly } y\right)
={}& \binom{L}{y}
\sum_{j=0}^{L-y} (-1)^j \binom{L-y}{j} \\
&\times
\frac{\left(2(m-j-y)-1\right)!!}{(2m-1)!!}.
\end{split}
\label{eq:prob_y_edges}
\end{equation}

Let $n=N/2$. Conditioned on the event $A_\nu \cap B_\nu$, the remaining graph is a uniformly random perfect matching on the remaining $N-4$ vertices. The desired bond pairs are precisely the $n-2$ singlets in the memory $\mu$ other than $(ik)$ and $(jl)$. Therefore,
\begin{widetext}
\begin{equation}
\begin{split}
    \Pr\! \left( Y_\nu \geq \frac{r}{2} \, \middle| \, A_\nu \cap B_\nu \right) &= \sum_{y=\frac{r}{2}}^{n-2} \binom{n-2}{y} \sum_{j=0}^{n-2-y} (-1)^j \binom{n-2-y}{j} \frac{\left(2(n-2-j-y)-1\right)!!}{(2\left(n-2)-1\right)!!} \\
    &= \sum_{y=\frac{r}{2}}^{n-2} \sum_{j=0}^{n-2-y} (-1)^j \, 2^{j+y} \, \frac{1}{y! \, j!} \, \frac{\left[(n-2)!\right]^2}{\left[2(n-2)\right]!} \, \frac{\left[2(n-2-j-y)\right]!}{\left[(n-2-j-y)!\right]^2}.
\end{split}
\label{eq:prob_y_r2_edges}
\end{equation}
\end{widetext}
The dominant contribution to the sum comes from $y,j=o(N)$, for which Stirling's approximation in the limit of large $N$ gives
\begin{equation}
\frac{\big[(n-2)!\big]^2}{\big[2(n-2)\big]!} \,
\frac{\big[2(n-2-j-y)\big]!}
     {\big[(n-2-j-y)!\big]^2}
\sim
2^{-2(j+y)}.
    \nonumber
\end{equation}
Therefore,
\begin{equation}
    \Pr\! \left( Y_\nu \geq \frac{r}{2} \, \middle| \, A_\nu \cap B_\nu \right) \sim  \sum_{y=\frac{r}{2}}^{n-2} \frac{1}{2^y \, y!} \sum_{j=0}^{n-2-y} \frac{(-1)^j}{2^j \, j!}.
\end{equation}
The inner sum converges to $e^{-1/2}$ as $n \to \infty$, uniformly for $y=o(n)$. We choose $r = c\log N$. The outer sum is dominated by its first term at $y=r/2$ since successive terms decrease by a factor $1/(2(y+1)) = O(1/\log N)$. Finally, we can write the probability as
\begin{equation}
    \Pr\! \left( Y_\nu \geq \frac{r}{2} \, \middle| \, A_\nu \cap B_\nu \right) \sim  e^{-1/2} \frac{(1/2)^{r/2}}{(r/2)!}.
    \label{eq:prob_y_r2_edges_simplified}
\end{equation}

An analogous analysis applies to $\Pr\! \left( Y_\nu \geq r \, \middle| \, A_\nu \oplus B_\nu \right)$. The only difference is that the total number of desired bond pairs is now $n-3$ rather than $n-2$. This is because, conditioned on exactly one of $(ik)$ and $(jl)$ belonging to the random memory $\nu$, two dimers of the memory $\mu$ must be broken instead of one. The same asymptotic analysis therefore yields
\begin{equation}
    \Pr\! \left( Y_\nu \geq r \, \middle| \, A_\nu \oplus B_\nu \right) \sim e^{-1/2} \frac{(1/2)^{r}}{r!}.
    \label{eq:prob_y_r_edges_simplified}
\end{equation}

Using Stirling's approximation on $r=c\log N$ and substituting Eqs.~(\ref{eq:prob_y_r2_edges_simplified}) and~(\ref{eq:prob_y_r_edges_simplified}) into Eq.~(\ref{eq:prob_r_nu_single}), we obtain
\begin{equation}
    \Pr \! \left(|X_\nu| \geq r = c\log N\right) \sim N^{-c \log \log N + O(1)}.
\end{equation}
Eq.~(\ref{eq:prob_r_nu_union_bound}) becomes
\begin{equation}
\Pr \!\left(\exists\,\nu\neq\mu:\,|X_\nu| \geq r\right) \leq (K-1) N^{-c \log \log N + O(1)}.
\label{eq:prob_r_nu_final}
\end{equation}
For any polynomial number of stored memories, $K=N^{O(1)}$, the right-hand side vanishes as $N\to\infty$. The contribution from the rare events is asymptotically negligible, and the first term in Eq.~(\ref{eq:prob_x_nu}) dominates. We now turn to its analysis.

Using Bernstein inequality, we can bound the first term in Eq.~(\ref{eq:prob_x_nu}) as follows
\begin{equation}
    \Pr\!\left(\sum_{\nu\neq\mu}\widetilde{X}_\nu \geq \frac{t}{2}\right) \leq \exp\!\left( - \frac{\frac{1}{2}\left(\frac{t}{2}\right)^2}{\sum_{\nu\neq\mu} \mathbb{E}\!\left[ \widetilde{X}_\nu^2 \right] + \frac{1}{3}\,r\,\frac{t}{2}} \right).
    \label{eq:berstein}
\end{equation}
It remains to estimate $\mathbb{E}\!\left[ \widetilde{X}_\nu^2 \right] = \mathbb{E}\!\left[ X_\nu^2 \right] - \mathbb{E}\!\left[ R_\nu^2 \right]$. Since
\begin{align}
\begin{split}
    \mathbb{E}\!\left[ R_\nu^2 \right] =
\mathbb{E}\!\left[X_\nu^2 \, \mathbf{1}_{|X_\nu|>r}\right] &\leq \left( \max X_\nu^2 \right)  \Pr \! \left(|X_\nu| \geq r\right) 
\end{split} \nonumber \\
\begin{split}
    &\leq \left( N - 4 \right)^2 N^{-c \log \log N + O(1)},
\end{split}
\end{align}
the contribution from the rare-event tail vanishes in the thermodynamic limit. From Eq.~(\ref{eq:var_x_nu}), we obtain
\begin{equation}
\mathbb{E}\!\left[ \widetilde{X}_\nu^2 \right] \sim \mathbb{E}\!\left[ X_\nu^2 \right] \sim \frac{3}{N}.
\end{equation}
Substituting this estimate into Eq.~(\ref{eq:berstein}) yields
\begin{equation}
    \Pr\!\left(\sum_{\nu\neq\mu}\widetilde{X}_\nu \geq \frac{t}{2}\right) \leq \exp\!\left( - \frac{\frac{1}{8}(N-4)^2}{\frac{3(K-1)}{N} + \frac{c(N-4)\log N}{6}} \right).
\end{equation}
When $K=O(N^2\log N)$, the second term in the denominator dominates, so the exponent scales with $N/\log N$, and the probability is exponentially suppressed. For larger memory loads, $K\gg N^2\log N$, the first term dominates, giving
\begin{equation}
    \Pr\!\left(\sum_{\nu\neq\mu}\widetilde{X}_\nu \geq \frac{t}{2}\right) \leq \exp\!\left[-O\!\left(\frac{N^3}{K}\right)\right].
    \label{eq:prob_x_tilde_nu_final}
\end{equation}
Combining Eqs.~(\ref{eq:prob_x_nu}), (\ref{eq:prob_r_nu_final}) and (\ref{eq:prob_x_tilde_nu_final}), and applying the union bound over the \[
\frac{N}{2}\left(\frac{N}{2}-1\right)=O(N^2)
\]
possible local bond rearrangements, we obtain
\begin{equation}
    \Pr \! \left( \mathrm{instability} \right) \leq \frac{N}{2} \left( \frac{N}{2} - 1\right) \exp\!\left[-O\!\left(N^3/K\right)\right].
    \label{eq:qhop_instability_prob_h2}
\end{equation}
The probability that a memory develops a local instability vanishes in the thermodynamic limit provided
\[
K_\text{max}^q=O\!\left(\frac{N^3}{\log N}\right),
\]
establishing the scaling of the storage capacity discussed in Sec.~\ref{section:local_stability_analysis}.
% establishing a scaling the storage capacity. This bound is conservative because the union bound neglects correlations between different local bond rearrangements and overestimates the probability of instability.

Finally, we note that the same analysis extends naturally to higher-order Hamiltonians. Take
\[
H_3^q = +\sum_\mu \sum_{(ij)<(kl)<(pq) \in \mu} \left(\boldsymbol{\sigma}_{i} \!\cdot\! \boldsymbol{\sigma}_{j}\right) \left(\boldsymbol{\sigma}_{k} \!\cdot\! \boldsymbol{\sigma}_{l}\right)\left(\boldsymbol{\sigma}_{p} \!\cdot\! \boldsymbol{\sigma}_{q}\right).
\]
as an example. The key difference is that, for a local bond swap, the selected memory now contributes $O(N^2)$ stabilizing terms instead of $O(N)$. The destabilizing contributions from the remaining memories have the same $O(1/N)$ expectation value and variance as before. The Bernstein inequality has the same denominator up to constants, while the numerator scales as $O(N^4)$ rather than $O(N^2)$. It follows that the probability of instability remains exponentially suppressed if $K_\text{max}^q=O\!\left(N^5/\log N\right)$. More generally, the $p$th-order Hamiltonian $H_p^q$ with $p \geq 2$ has a capacity scaling as $K_\text{max}^q=O\!\left(N^{2p-1}/\log N\right)$.

\section{Capacity of Classical Hopfield Model}
\label{section:capacity_classical_hopfield_appendix}
The same local stability analysis can be carried out for the classical Hopfield model. The second-order Hamiltonian is
\begin{equation}
    H_2^c
    =
    -\sum_\mu^K \sum_{i<j}^N
    \xi_i^\mu \xi_j^\mu
    \sigma_i^z \sigma_j^z,
\end{equation}
where the stored memories are the product states $\ket{\psi_c^\mu}=\bigotimes_{i=1}^N \ket{\xi_i^\mu}$ with $\xi_i^\mu=\pm1$. Since the memories are orthogonal eigenstates of the Hamiltonian, it is straightforward to show that the only relevant local errors are single-spin flips generated by $\sigma^x_i$. As in the quantum dimer model, we determine the storage capacity by evaluating the local energy change of the memories under these errors,
\begin{equation}
\begin{split}
    \Delta E_k &= \bra{\{\xi_i^\nu \}} \sigma_k^x H_2^c \sigma_k^x \ket{\{\xi_i^\nu \}} - \bra{\{\xi_i^\nu \}} H_2^c \ket{\{\xi_i^\nu \}} \\
    &= 2(N-1) + 2\sum_{\mu \neq \nu} \sum_{i \neq k} \xi^\mu_i \xi^\mu_k \xi^\nu_i \xi^\nu_k \\
    &\equiv S + R,
\end{split}
\end{equation}
where $S=2(N-1)$
is the signal arising from the target memory $\nu$, while
\[R=2\sum_{\mu\neq\nu}\sum_{i\neq k}\xi_i^\mu\xi_k^\mu\xi_i^\nu\xi_k^\nu\]
is the crosstalk noise from the other stored memories.
\begin{figure}
    \centering
    \subfloat[]{
    \includegraphics[width=0.47\linewidth]{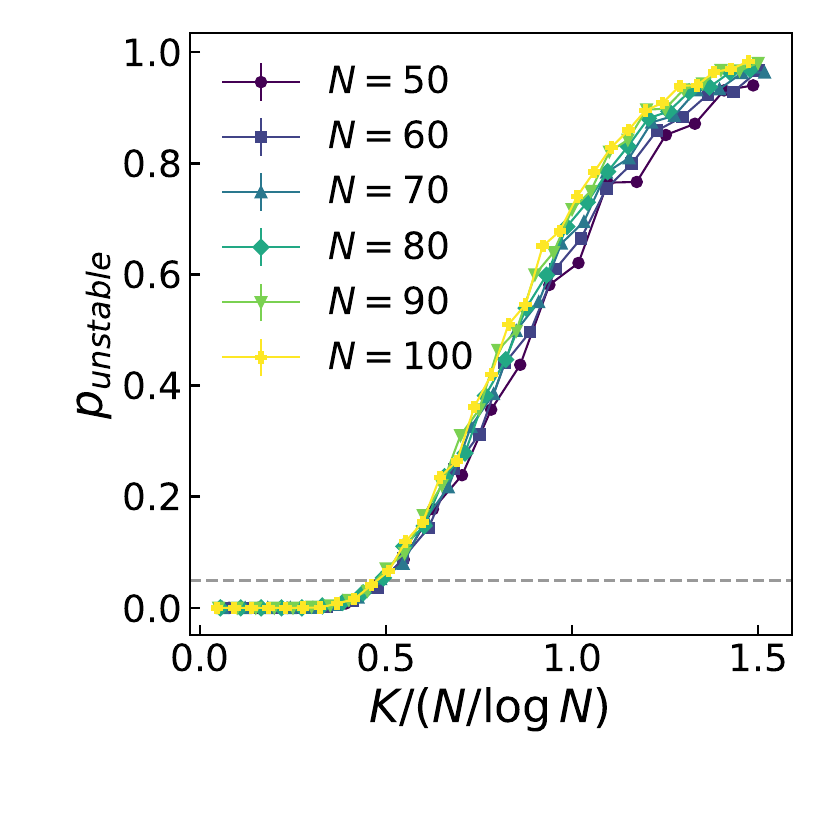}
    \label{fig:chop_error_prob_scaling_h2}}
    \hfill
    \subfloat[]{
    \includegraphics[width=0.47\linewidth]{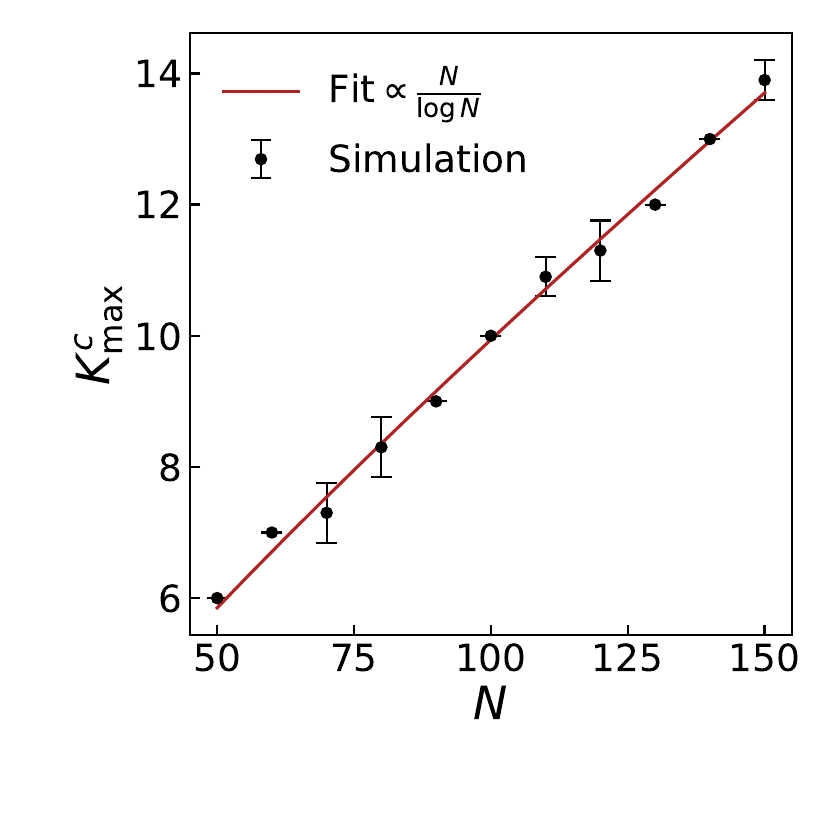}
    \label{fig:chop_capacity_h2}}
    \caption{(a) The probability of instability as a function of the rescaled memory load for different system sizes. The dashed gray line indicates the threshold $p_0$ used to define the finite-size storage capacity $K_\text{max}^c$. (b) $K_\text{max}^c$, defined as the maximum number of stored memories for which the instability probability is below $p_0=5\%$, as a function of $N$. The numerical results are consistent with the predicted scaling $K_\text{max}^c = O(N/\log N)$. See Sec.~\ref{section:local_stability_analysis} in the main text for detail on how the simulations are done.}
    \label{fig:chop_local_stability_h2}
\end{figure}
In the thermodynamic limit, the noise term $R$ can be approximated by a Gaussian random variable,
\[
R \sim \mathcal{N}\!\left(0,\,4(K-1)(N-1)\right),
\]
by the CLT. The probability that a single-spin flip lowers the energy is therefore
\begin{equation}
    \Pr \! \left( R < -2(N-1) \right) = 1 - \Phi \!\left( \sqrt{\frac{N-1}{K-1}} \right),
    \label{eq:chop_prob_r}
\end{equation}
where 
\[\Phi (x) = \frac{1}{2} + \frac{1}{2} \operatorname{erf}\!\left( \frac{x}{\sqrt{2}} \right) \]
is the cumulative distribution function of the standard normal distribution. Assume the storage capacity $K_c = \alpha N$ in the large $N$ limit with $\alpha \ll 1$. Then
\[
\sqrt{\frac{N-1}{K_c-1}}
\sim
\frac{1}{\sqrt{\alpha}}
\gg1,
\]
allowing us to use the asymptotic expansion
\[
\operatorname{erf}\!\left(\frac{x}{\sqrt{2}}\right)
\approx
1-\sqrt{\frac{2}{\pi}}\,
\frac{e^{-x^2/2}}{x},
\quad x\gg1.
\]
Eq.~(\ref{eq:chop_prob_r}) then becomes
\begin{equation}
    \Pr \! \left( R < -2(N-1) \right) = \sqrt{\frac{\alpha}{2\pi}} e^{-\frac{1}{2\alpha}}.
\end{equation}
Applying the union bound over all $N$ independent single-spin flips, we obtain
\begin{equation}
    \Pr \! \left( \mathrm{instability} \right) \leq N \sqrt{\frac{\alpha}{2\pi}} e^{-\frac{1}{2\alpha}}.
\end{equation}
Requiring this probability to vanish in the TDL implies $\alpha=O\!\left(1/\log N\right)$. Hence, the storage capacity scales as
\[
K_\text{max}^c=O\!\left(\frac{N}{\log N}\right).
\]
Plotted in Fig.~\ref{fig:chop_local_stability_h2} are (a) the instability probability and (b) the finite-size capacity. Both are consistent with the predicted scaling $K_\text{max}^c = O(N/\log N)$.

\section{Mattis Overlap in Classical Hopfield Models}
\label{section:mattis_classical_hopfield}
In this Appendix, we show that the maximum absolute Mattis overlap,
\begin{equation}
\label{eq:chop_mattis_abs}
    m_{\max}^c= \max_\mu \, \left| m_\mu^c \right| = \max_\mu \, \left\langle \left| \frac{1}{N} \sum_i \xi_i^\mu \sigma_i^z \right| \right\rangle,
\end{equation}
evaluated in the ground state of the second-order classical Ising Hamiltonian
\begin{equation}
H_2^c = -\sum_\mu \sum_{i<j} \xi^\mu_i \xi^\mu_j \sigma_i^z \sigma_j^z,
% \label{eq:chop_h234}
% \begin{split}
    % H_2^c =& -\sum_\mu \sum_{i<j} \xi^\mu_i \xi^\mu_j \sigma_i^z \sigma_j^z, \\
    % H_3^c =& -\sum_\mu \sum_{i<j<l} \xi^\mu_i \xi^\mu_j \xi^\mu_l \sigma_i^z \sigma_j^z \sigma_l^z,  \\
    % H_4^c =& -\sum_\mu \sum_{i<j<l<k} \xi^\mu_i \xi^\mu_j \xi^\mu_l \xi^\mu_k \sigma_i^z \sigma_j^z \sigma_l^z \sigma_k^z,  \\
% \end{split}
\end{equation}
exhibits the same qualitative features as the quantum singlet Mattis overlap discussed in Sec.~\ref{section:mattis}.
% exhibits the expected finite-size scaling collapse after accounting for extreme-value statistics. Specifically, we will see that for the $p_c$-th-order Hamiltonian $H_{p_c}^c$, the storage capacity scales as $O(N^{p_c-1})$. 

% \begin{figure}
%     \centering
%     \subfloat[]{
%     \includegraphics[width=0.47\linewidth]{chop_gs_rescaled_mattis_overlap_scaling_h2.pdf}
%     \label{fig:chop_gs_mattis_overlap_h2}}
%     \hfill
%     \subfloat[]{
%     \includegraphics[width=0.47\linewidth]{chop_gs_rescaled_mattis_overlap_scaling_h3.pdf}
%     \label{fig:chop_gs_mattis_overlap_h3}}
%     \hfill
%     \subfloat[]{
%     \includegraphics[width=0.47\linewidth]{chop_gs_rescaled_mattis_overlap_scaling_h4.pdf}
%     \label{fig:chop_gs_mattis_overlap_h4}}
%     \caption{The rescaled maximum Mattis overlap (Eq.~\ref{eq:rescaled_mattis_c}) of the ground state of the (a) second-, (b) third- and (c) fourth-order Hamiltonians defined in Eq.~\ref{eq:chop_h234}. The data collapse in the non-retrieval phase agrees with the power-law scaling of the storage capacity predicted by the analytical approaches.}
%     \label{fig:chop_gs_mattis_overlap}
% \end{figure}

\begin{figure}
    \centering
    \includegraphics[width=0.9\linewidth]{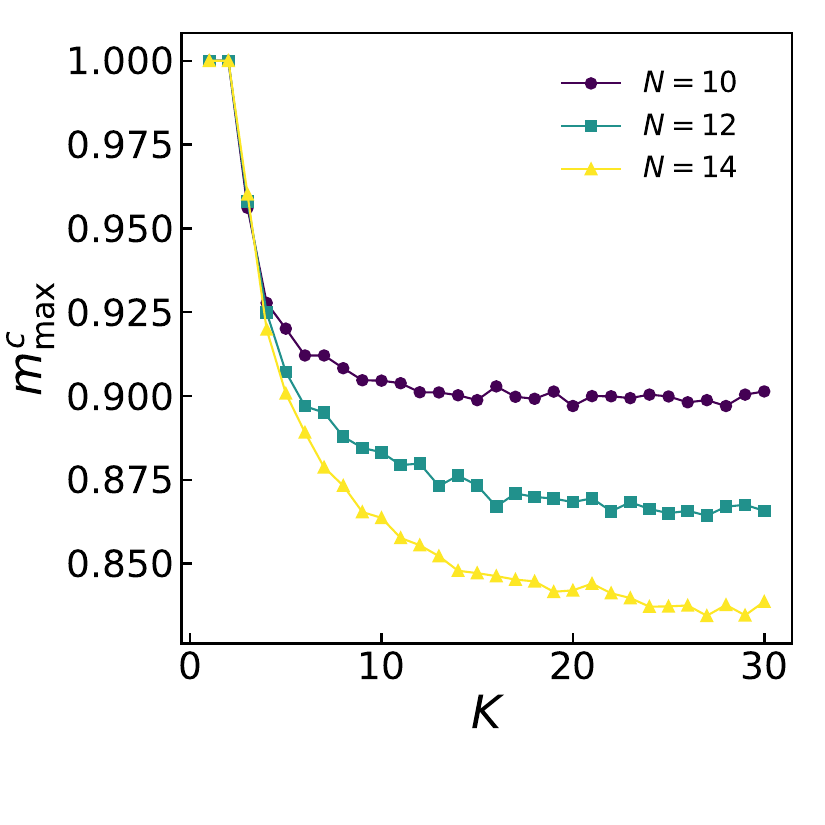}
    \caption{Maximum absolute Mattis overlap (Eq.~\ref{eq:chop_mattis_abs}) as a function of memory loading $K$ for different system sizes $N$. The overlap remains finite even at large memory loads.}
    \label{fig:chop_gs_mattis_overlap}
\end{figure}

We consider the absolute value because the even-order Hamiltonians possess a global $\mathbb{Z}_2$ symmetry, under which a pattern and its antipattern are degenerate. The ground state obtained numerically may be either a pattern or its antipattern. 

Plotted in Figure~\ref{fig:chop_gs_mattis_overlap} is the maximum absolute Mattis overlap. A crossover between the retrieval and non-retrieval regimes is visible between $K=2$ and $K=3$. For the finite system sizes considered here, the expected linear growth of the storage capacity with $N$ produces only a modest shift in this crossover, making the finite-size trend difficult to resolve. Deep in the non-retrieval regime, the maximum Mattis overlap remains finite. The ground state can retain a substantial overlap with one of the stored patterns even after the memories have lost their local stability.

\bibliography{apssamp}

\end{document}